%% file: W4OVAW.tex
\documentclass[journal]{vgtc}                     % final (journal style)

\usepackage{amsmath}
\usepackage{amssymb}

\usepackage{comment}

\onlineid{0}

\vgtccategory{Research}

\title{Optimizing Visual Analytics Workflows: From Theory to Practice}

\author{%
  \authororcid{Philip\ Beaucamp}{0009-0007-2106-9516},
  \authororcid{Alfie\ Abdul-Rahman}{0000-0002-6257-876X},
  \authororcid{Rita\ Borgo}{0000-0003-2875-6793},
  \authororcid{Wolfgang\ Jentner}{0000-0003-1045-6020},
  \texorpdfstring{\\}{, }
  \authororcid{Saiful\ Khan}{0000-0002-6796-5670},
  \authororcid{Yiwen\ Xing}{0000-0003-1521-6616},
  \authororcid{David\ Ebert}{0000-0001-6177-1296},
  and 
  \authororcid{Min\ Chen}{0000-0001-5320-5729}
}

\authorfooter{
  \item
  P. Beaucamp, Y. Xing, and M. Chen are with University of Oxford, UK. E-mail: \{philip.beaucamp, yiwen.xing, min.chen\}@eng.ox.ac.uk
  \item
  A. Abdul-Rahman and R. Borgo are with King's College London, UK. E-mail: \{alfie.abdulrahman, rita.borgo\}@kcl.ac.uk.
  \item W. Jentner and D. Ebert are with University of Arizona, USA. E-mail: \{wjentner, ebertd\}@arizona.edu.
  \item S. Khan is with the Science and Technology Facilities Council, UK. E-mail: saiful.khan@stfc.ac.uk.
}

\abstract{%
  The principle of visual analytics (VA) is to provide integrated workflows where human-centric processes (e.g., visualization and interaction) and machine-centric processes (e.g., statistics and algorithms) complement each other. To implement this principle in practice, it is necessary to reason about the trade-offs among different processes and make optimal use of them in a workflow.
  Building on an existing ontology of the methodology for analyzing such trade-offs information-theoretically and for optimizing VA workflows systematically, we investigate ways to transform this methodology from theory to practice. In particular, we adopted the action research method. Through case studies in different application domains, VA researchers with different background knowledge and experiences offered their answers to several hypotheses about using the methodology in practice and proposed ways forward. In this paper, we present our collective analysis, the strengths and feasibility of this theory-based methodology, as well as the obstacles to its broad deployment in practice. To address these challenges, we outline a roadmap to remove such obstacles.
}


\graphicspath{{figs/}{figures/}{pictures/}{images/}{./}} % where to search for the images

\usepackage{tabu}                      % only used for the table example
\usepackage{booktabs}                  % only used for the table example
\usepackage{lipsum}                    % used to generate placeholder text
\usepackage{mwe}                       % used to generate placeholder figures
\usepackage{ccicons}                   % package to be able to use icons from creative commons

\usepackage{mathptmx}                  % use matching math font

\usepackage{booktabs}
\usepackage{tabularx}
\usepackage{array}

\begin{document}

%%%%%%%%%%%%%%%%%%%%%%%%%%%%%%%%%%%%%%%%%%%%%%%%%%%%%%%%%%%%%%%%
%%%%%%%%%%%%%%%%%%%%%% START OF THE PAPER %%%%%%%%%%%%%%%%%%%%%%
%%%%%%%%%%%%%%%%%%%%%%%%%%%%%%%%%%%%%%%%%%%%%%%%%%%%%%%%%%%%%%%%


\input{Sections/1.Introduction}
\input{Sections/2.RelatedWork}
\input{Sections/3.Problem}
\input{Sections/4.CaseStudies}

\input{Sections/5.HypotheseAnalysis}

\input{Sections/6.W4OVAW}
\input{Sections/7.Discussions}

\input{Sections/8.Conclusions}
\acknowledgments{
    This work is part of the VIS4ML4HD project funded by UK Research and Innovation (EPSRC: EP/X029557/1).
    Some co-authors used \emph{Grammarly} to check grammar and spelling, and for occasional rephrasing. 
}

\bibliographystyle{abbrv-doi-hyperref}

\bibliography{W4OVAW}

\appendix % You can use the `hideappendix` class option to skip everything after \appendix
\crefalias{section}{appendix} % this is to make sure that cleverref switches to referring to Appx. X from here on

\clearpage

\begin{center}
\large
APPENDICES\\[1mm]
\Large \noindent
\textbf{\textsf{Optimizing Visual Analytics Workflows:\\
 From Theory to Practice}}\\[2mm]
\normalsize
P.\ Beaucamp$^1$, A.\ Abdul-Rahman$^2$, R.\ Borgo$^2$, W.\ Jentner$^4$,\\[1mm]
 S.\ Khan$^3$, Y.\ Xing$^1$, D.\ Ebert$^4$, M.\ Chen$^1$\\[1mm]
$^1$University of Oxford, UK\\
$^2$King's College London, UK\\
$^3$Science and Technology Facilities Council (STFC), UK\\
$^4$University of Arizona, USA
\normalsize
\end{center}

This document consists of eight appendices that support the main body of the paper. The first seven appendices are reports on the seven case studies by individual participants of this action research project. Each report mainly contains the original text produced during the action research, with relatively minor updates incorporated during the writing of the main body of the paper. More dynamically-updated versions of these case studies can be found on the IVAS web site \cite{IVAS:2026:web}. In addition, we characterized the case study reports on the IVAS web site \cite{IVAS:2026:web} using the observational attributes discussed in Section \ref{sec:CS-Summary}. Characterizations are summarized in the Appendix \ref{appendix:H}.
These eight appendices are:

\begin{itemize}
    \item Appendix \ref{appendix:A ML4Finance}: \textbf{CS\#1 ML4Finance}
    -- A retrospective analysis of how a VA workflow in finance for mutual fund selection was improved \cite{Yan2025FundSelector:Selection,Yan2025VisualAnalysis}.
    \item Appendix \ref{appendix:B SimilarityDetection}: \textbf{CS\#2 SimilarityDetection} -- A retrospective analysis of how a VA workflow in the digital humanities for detecting similar texts in the literature of the 18th century was improved \cite{abdulrahman:2017:CGF}.
    \item Appendix \ref{appendix:C BookTrade}: \textbf{CS\#3 BookTrade} -- A retrospective analysis of how a VA workflow in the digital humanities for visualizing the data of early book trade was improved \cite{xing:2024:TVCG}.
    \item Appendix \ref{appendix:D DataVirtualization}: \textbf{CS\#4 DataVirtualization} -- A primarily prospective analysis of how a traditional ML workflow for data and learning preparation could be improved. The analysis guided the infrastructure design and development planning. Parts of the infrastructure have been implemented \cite{Khan:2025:SCC}, while the other parts are being implemented.   
    \item Appendix \ref{appendix:E SubspaceAnalysis}: \textbf{CS\#5 SubspaceAnalysis} -- A primarily retrospective analysis of how a traditional workflow in data mining for subspace analysis was improved \cite{jentner2023}. The report includes some prospective analysis about how the VA workflow could be further improved.
    \item Appendix \ref{appendix:F Prompts4LLMs}: \textbf{CS\#6 Prompts4LLMs} -- A retrospective analysis of how a VA workflow in finance for annotating visualization plots was improved \cite{Hao2025}.
    \item Appendix \ref{appendix:G GlacierMovement}: \textbf{CS\#7 GlacierMovement} -- A retrospective analysis of how the visualization capability of a workflow in geoscience was improved with the development of a new visual design and supporting algorithms \cite{borgo-2010}.
    \item Appendix \ref{appendix:H}: A summary of existing case studies conducted before 2025 \cite{IVAS:2026:web}, i.e., before this project.
\end{itemize}

\section{Case Study 1 - VA + ML for Mutual Fund Selection
(ML4Finance)}
\label{appendix:A ML4Finance}
\input{Appendix/1.CaseStudyVA+ML}
\section{Case Study 2 - Text Similarity Detection}
\label{appendix:B SimilarityDetection}
\input{Appendix/2.CaseStudy-TextSimilarityDetection}

\section{Case Study 3 - Book Trade Visualization}
\label{appendix:C BookTrade}
\input{Appendix/3.CaseStudy3BookTrade}
\section{Case Study 4 - Data Virtualization for Machine Learning}
\label{appendix:D DataVirtualization}
\input{Appendix/4.CaseStudy4-DataVirtualization}
\section{Case Study 5 - Multi-Dimensional Pattern Exploration (SubspaceAnalysis)}
\label{appendix:E SubspaceAnalysis}
\input{Appendix/5.CaseStudy}
\section{Case Study 6 - Prompts4LLMs}
\label{appendix:F Prompts4LLMs}
\input{Appendix/6.CaseStudyPromptsforLLMs}
\section{Case Study 7 - Spatio-Temporal Visualization of Glacier Terminus Movements (GlacierMovement)}
\label{appendix:G GlacierMovement}
\input{Appendix/7.CaseStudy}

\input{Appendix/AppendixH}

\end{document}

%% file: Sections/1.Introduction.tex
\firstsection{Introduction\label{sec:introduction}}
\maketitle
%% \section{Introduction} %for TVCG use above \firstsection{..} instead

According to Jim Thomas and his colleagues who coined the concept of \emph{visual analytics} (VA), VA ``\emph{is the science of analytical reasoning facilitated by interactive visual interfaces}'' \cite{Wong:2004:CGA,Thomas:2005:book}. Keim et al. further defined VA as a technique that ``\emph{combines automated analysis techniques with interactive
visualizations for an effective understanding}'' of, ``\emph{reasoning}'' about, ``\emph{and decision making}'' based on very large and complex datasets \cite{Keim:2008:book}. During the past two decades, many VA systems and tools have been developed for a variety of applications, ranging from physical sciences to social sciences, from finance to digital humanities, and from public safety to sports.

Today, designing, evaluating, and improving VA systems is a primary area of activity in our discipline. VA researchers and practitioners are commonly relying on \emph{user-centered requirement analysis and evaluation} \cite{Sedlmair:2012:TVCG}. Although such processes appeared in many VA papers as ``waterfall'' processes (i.e., requirements first and evaluation at the end), many VA researchers and practitioners adopted incremental, iterative, nested, agile, and evolutionary development approaches in practice \cite{Larman:2003:book,Munzner:2009:TVCG,Martin:2014:book}. 

One may easily notice that the user-centered design approach is often not the dominant approach in many disciplines, which are relatively more mature than visualization. For example, bridge designers and building architects rely extensively on their technical knowledge, methods, tools, and skills to analyze the requirements and constraints in many aspects of their projects, of which the end users usually have little knowledge. A physician may consult other physicians about a specific diagnosis and treatment plan, rather than responding to the feedback of random participants through a user study. The heavy reliance on the user-centered approach in the visualization discipline reflects a lack of auditable, effective, and efficient methods in the discipline \cite{Chen:2025:CGA}. Although it is, and will always be, necessary to engage with end-users in both requirements analysis and evaluation, over-emphasizing or over-relying on the user-centered approach may discourage design initiatives of VA researchers and practitioners and impede the development of new auditable, effective, and efficient methods. 

Chen and Ebert made an attempt to introduce a systematic methodology \cite{Chen:2019:CGF} for VA system designers to optimize a VA system in a structured and iterative manner, namely (a) identifying problems (metaphorically termed \emph{symptoms}) in different processes of a VA workflow, considering possible \emph{causes}, deliberating optional \emph{remedies}, determining the optimal solution and analyzing potential \emph{side-effects}. The side-effects are then considered as new symptoms, leading to a new iteration.

The methodology was formulated based on the theoretical development of information-theoretic cost-benefit analysis \cite{Chen:2016:TVCG}. 
It has been applied to several case studies \emph{retrospectively} (e.g., \cite{Zhao:2017:THS,Zhang:2016:CGF,Zhang:2018:CHI} reported in \cite{Chen:2019:CGF}) and to several other case studies \emph{prospectively} (e.g., \cite{Ye:2023:TVCG,Jin:2024:TVCG}).
In addition, a few other retrospective and prospective case studies were reported on the website created for the methodology \cite{IVAS:2026:web}. However, the knowledge and experience of this methodology had not been extended before this work, suggesting that there must be some obstacles that hinder its broad adoption and application.

To investigate possible obstacles, we adopted the qualitative research method, \emph{Action Research} \cite{Reason:2001:book,McNiff:2013:book}, which is an interactive and collaborative inquiry process where researchers/participants investigate a problem while taking actions to solve the problem. Six researchers, who have a different amount of VA experience but no prior experience in using the methodology, conducted seven case studies. Together with two researchers with prior experience, they analyzed the actions taken in these case studies and identified several ways to improve the methodology to make it more usable in practice.

Our contributions include:
\begin{itemize}
    \item We conducted \emph{Action Research} to investigate a challenging research problem, i.e., how to transform a theoretical methodology for optimizing VA workflows into a practical one.%
    \item Our case studies, as the action part, demonstrate that the methodology can be applied to a variety of VA workflows, e.g., different applications, different amounts of visualization (e.g., integrated VA, visual design), and retrospective or prospective.%
    \item The analysis of our actions results in several conceptual improvements of the methodology, including a new \emph{methodological workflow} for VA practitioners to optimize VA workflows, and an analysis of different metaphors for assisting the understanding of the information-theoretical concept that underpins the methodology.%
    \item The analysis of our actions provides a roadmap for further technical development, including a repository of case studies and an interactive tool to assist VA practitioners in using the methodology in practice.  
\end{itemize}

%% file: Sections/2.RelatedWork.tex
\section{Related Work}
\label{sec:RelatedWork}
%
% ----------
\noindent
\textbf{Abbreviations.}
In this section and in all following sections, we refer to the methodology of Chen and Ebert \cite{Chen:2019:CGF} as \textbf{SCORE} (Symptom, Cause, Optimize, Remedy, side-Effect). Other abbreviations used include:
\textbf{AR} (Action Research),
\textbf{ML} (Machine Learning),
\textbf{VA} (Visual Analytics),
\textbf{VIS} (Visualization and VA), and
\textbf{VIS4ML} (VIS for ML).

% ----------
\vspace{2mm}\noindent
\textbf{Action Research (AR).}
AR is an established methodology in the social sciences that integrates knowledge generation with practical intervention. Lewin’s early formulation~\cite{lewin1946} frames AR as a form of comparative research on social action that proceeds through a ``spiral of steps'' involving planning, action, and fact-finding about the effects of an intervention. This positions knowledge production and interventions as mutually informed processes. Today, the scope of AR is much more broad \cite{Reason:2001:book,McNiff:2013:book}. For example, Hayes characterized AR as being (i) democratic, (ii) collaborative, (iii) context-sensitive, (iv) with inquiry conducted with people experiencing real problems, (v) a focus on highly contextualized and localized solutions, and (vi) open-ended, iterative cycles of action and reflection~\cite{hayes2011}. Hayes considered that AR provides a rigorous methodology for socially engaged research.
Sein et al. applied AR to information systems research to design IT artifacts and proposed a method called Action Design Research~\cite{Sein2011}.
McCurdy et al. showed that AR principles can be productively translated into visualization design research, through attention to intervention, reflection, reciprocal shaping, and evaluation~\cite{McCurdy2016}.

In this work, we adopt the AR methodology to study the research problem outlined in Section~\ref{sec:introduction}. In particular, VIS researchers with different backgrounds work collaboratively to formulate hypotheses, design and conduct interventions, observe and analyze outcomes and experiences, and propose new solutions.

%one that is carried out with participants rather than on them and that privileges contextualized problem-solving, iterative intervention, and critical reflection. 

% This cyclical and reflective characteristics make AR especially relevant to research areas such as visualization and visual analytics, where situated deployment, stakeholder collaboration, and progressive refinement are central to both design and development processes.

% Further developments include Sein et al.’s Action Design Research (ADR)~\cite{Sein2011}, which builds on AR, adapting it to information systems research. ADR emphasizes the concurrent interweaving of building, intervention, and evaluation in the development of artefacts. McCurdy et al.~\cite{McCurdy2016} show how action-research principles can be productively translated into visualisation design research, through attention to intervention, reflection, reciprocal shaping, and evaluation in context.
 %To our best knowledge, the action research method has not yet been reported in the VIS literature.

% ----------
\vspace{2mm}\noindent
\textbf{Qualitative Research in VIS.}
In VIS research, the role of qualitative research methods has become increasingly salient, facilitating the scientific, rigorous, efficient, and effective exploration of theoretical and practical solutions.  
%shifting from a foundational function in assembling requirements to a well-established, meaningful core contribution.
%
Munzner proposed a nested model for developing visualization solutions, which comprises four layers: domain problem characterization, operations and data abstraction, interaction and visual encoding, and algorithm development~\cite{Munzner:2009:TVCG}.
Sedlmair et al. outlined a nine-stage methodology to carry out design studies \cite{Sedlmair:2012:TVCG}.
Multiple scholars built on this methodology and provided  extensions and variants, e.g., an interpretivist approach and six rigor criteria \cite{Meyer2019}, design by immersion with a varied mixture of design activities \cite{Hall2020}, and investigation of different perspectives of collaborative stakeholders in VIS design studies \cite{Akbaba2023}.
%
% present a nine-stage methodological framework for undertaking a design study, drawing on their expertise in conducting twenty-one design studies and reviewing many more, thereby establishing the design study as a central type of problem-driven research utilising qualitative methods.
%
% Transcending the field of design studies, qualitative VIS research has broadened significantly.
%
 Lam et al. conducted secondary research on VIS evaluation activities and categorized them into seven scenarios \cite{Lam2012}. Diehl et al. compared grounded theory workflows with visual analytics workflows, arguing that theoretical advances in VIS can benefit from grounded theory, while VIS tools can benefit research based on grounded theory \cite{Diehl2025}. Akbaba et al. discussed the impact of feminist epistemology on VIS research \cite{Akbaba2025}.

 One can find many forms of qualitative research in the VIS field, e.g., \emph{grounded theory} \cite{Diehl2025}, \emph{ethnography} \cite{Hall2020}, \emph{phenomenological research} \cite{Akbaba2023}, \emph{narrative research} \cite{Hullman:2011:TVCG}, \emph{action research} \cite{McCurdy2016}, and a variety of data collection methods such as interviews \cite{Kandel2012,Zhang2022}, focus groups \cite{Araujo:2025:IV}, and secondary research \cite{Lam2012}). Although this work is part of this broad scope and trend, it makes a novel attempt to transform a theoretical methodology to a practice using AR.
 
 %Furthermore, interview studies have become more prominent as independent inquiries. As an example, Kandel et al.\cite{Kandel2012} interviewed 35 data analysts in a semi-structured manner about enterprise data analysis workflows, while Zhang et al.\cite{Zhang2022} conducted a qualitative interview study of COVID-19 dashboard creators analyzing the processional cycle of design practices.
 
 % Nevertheless, qualitative research in VIS has hardly ever evaluated or enhanced theoretical methods supporting VA workflow design and analysis through case studies in which domain experts collectively assess a method’s practical viability. Therefore, our work will also help enrich qualitative research in the field of VIS.

\vspace{2mm}\noindent
\textbf{Information Theoretic Cost-benefit Analysis.}
Chen and J\"{a}nicke identified a range of visualization phenomena that could be explained using information theory, including overview-and-detail, logarithmic plot, redundancy, and so on \cite{Chen:2010:TVCG}. In particular, they showed that interactive visualization broke the conditions of \emph{data processing inequality} that is a major hindrance in automated data processing. Although this mathematically proves the value of visualization and interaction, it does not mean that interactive visualization is always superior to statistics, algorithms, or even reading data.   
Chen and Golan proposed an information-theoretic cost-benefit ratio to analyze trade-offs among different processes in data intelligence workflows \cite{Chen:2016:TVCG}.
Chen et al. further improved the mathematical definition of the cost-benefit ratio, while showing the effects of human knowledge and biases through both empirical data and information-theoretic measures \cite{Chen:2022:E1,Chen:2022:E2}.

Recognizing the difficulties in computing information-theoretic measures in VA processes at the moment, Chen and Ebert proposed a qualitative methodology, i.e., SCORE, for optimizing VA workflows \cite{Chen:2019:CGF}. Although guided by the above theoretical development, SCORE relies mainly on the abstract assessment of statistics, algorithms, visualization, and interaction in VA workflows, through the lens of symptoms, causes, remedies, and side-effects.
Ye and Chen used SCORE to improve an ML workflow by introducing a new variant of theme-river plots \cite{Ye:2023:TVCG}.
Jin et al. used it to address design issues in two tree plots (icicle and sunburst trees) by introducing the radial icicle tree (RIT) \cite{Jin:2024:TVCG}.
Saner and Chen used it to address data sparsity in the detection of similarity in music by transferring music data to visual data in an ML workflow \cite{Saner:2025:AS}.
In addition to these published works, there are also online reports of its application \cite{IVAS:2026:web}, including retrospective and prospective analysis of VA workflows \cite{Zhang:2016:CGF,Zhao:2017:THS,Rydow:2023:TVCG}, visual designs \cite{Zhang:2018:CHI}, and dashboards \cite{Bach:2023:TVCG}.

%% file: Sections/3.Problem.tex
\section{Research Questions and Methods}
\label{sec:Methods}
\textbf{Research Questions.} With information-theoretical abstraction, every process in a VA workflow can be considered a \emph{transformation} from an input data space to an output data space. In information theory, a data space is referred to as an \emph{alphabet} and a unique data point in the space is a \emph{letter} of the alphabet \cite{Chen:2016:TVCG}. For each process (which can be a composite process), there are three fundamental measures, which define the cost-benefit ratio of a transformation (i.e., process) as follows: 
\[
    \frac{\text{Benefit}}{\text{Cost}} =
    \frac{\text{Alphabet Compression (AC)} - \text{Potential Distortion (PD)}}{\text{Cost (Ct)}}
\]
The quantitative version of the formula and detailed description can be found in \cite{Chen:2016:TVCG}. Chen and Ebert broadly categorized VA processes into four groups, \emph{statistics} (stat), \emph{algorithms} (alg) (including ML models), \emph{visualization} (vis) and \emph{interaction} (int) \cite{Chen:2019:CGF}. Since these processes are many-to-one mappings, they can lead to information loss (i.e., AC > 0) and can cause errors (i.e., PD > 0), Meanwhile, smaller data spaces (e.g., aggregated time series) usually cost less to analyze (statistically or algorithmically), visualize, and interact with, while errors can result in adversarial cost. Therefore, they proposed optimizing VA workflows by reasoning about trade-offs among abstract entities [AC, PD, Ct]  $\times$ [stat, alg, vis, int]. Furthermore, they outlined a systematic approach to reason about ``symptoms'', ``causes'', ``remedies'', and ``side-effects'' using abstract entities, guiding the search for potential solutions to improve VA workflows. Hence, this methodology is referred to as SCORE (Symptom, Cause, Optimize, Remedy, side-Effect). However, the adoption of SCORE has so far been limited to several published papers and the original authors. This leads to the following hypotheses:% 

\begin{itemize}
    \item[\textbf{H1.}] A practitioner has to know information theory.\\
        $\hookrightarrow$ \textbf{Q1:} If $\textbf{H1}$ is true, how much does one need to know?
    \item[\textbf{H2.}] A practitioner has to know some psychology.\\
        $\hookrightarrow$ \textbf{Q2:} If $\textbf{H2}$ is true, how much does one need to know?
    \item[\textbf{H3.}] The method could be better described.\\
        $\hookrightarrow$ \textbf{Q3:} If $\textbf{H3}$ is true, what would be a better description?
    \item[\textbf{H4.}] The method needs additional procedures, e.g., for retrospective analysis, prospective analysis, learning, or experimenting.\\
        $\hookrightarrow$ \textbf{Q4:} If $\textbf{H4}$ is true, what could be the proposed new procedures?
    \item[\textbf{H5.}] There are other things that we can do to \emph{transform this largely theoretical method (to most people) to a practical method}.\\
        $\hookrightarrow$ \textbf{Q5:} If $\textbf{H5}$ is true, what are these other things?
\end{itemize}

\vspace{2mm}\noindent
\textbf{Research Method.} On the one hand, those who are not familiar with the SCORE methodology cannot offer meaningful answers to the above research questions. On the other hand, the original authors may be biased by their knowledge and experience and might not recognize obstacles easily. To address this dilemma, we adopted \emph{Action Research}, which allows \emph{participants} to learn by doing and \emph{researchers} to advance understanding by observing and analyzing actions \cite{Reason:2001:book,McNiff:2013:book}.   

Our project team consists of six participants who took \textbf{actions} and two original authors. All eight members analyzed the actions collectively as researchers. The six participants have different levels of VIS experience (ranging between 2 to 30+ years), but did not use SCORE until the project. As the \textbf{actions} of the project, they learned to use SCORE, and conducted seven case studies independently (one each except one participant conducted two). In each case study, a participant applied SCORE to analyze how a practical VA workflow was or could be optimized. In most case studies, the application is retrospective, except for two case studies that feature prospective analysis. We will briefly summarize these case studies in the next section, while providing more detailed case study reports in seven Appendices A--G.

After applying SCORE to a case study independently, the corresponding participant presented the case study to the whole team, while offering a personal evaluation of the aforementioned hypotheses. Participants used a Google folder to share their slides, case study reports, and relevant papers. Through seven online meetings and many email correspondences, the team collectively discussed and analyzed the case studies, evaluated the aforementioned hypotheses (Section \ref{sec:HypothesesEvaluation}), developed a new workflow to apply SCORE (Section \ref{sec:W4OVAW}), examined several metaphors that may help to appreciate the trade-offs among [AC, PD, Ct]  $\times$ [stat, alg, vis, int] and proposed a roadmap that includes a software solution (Section \ref{sec:Discussions}).

% Designing, evaluating, and improving visual analytics (VA) systems is a primary area of activities in our discipline. In this paper, we present an ontological framework for recording and categorizing technical shortcomings to be addressed in a VA workflow, reasoning about the causes of such problems, identifying technical solutions, and anticipating secondary effects of the solutions. The methodology is built on the theoretical premise that designing a VA workflow is an optimization of the costbenefit ratio of the processes in the workflow. It makes uses three fundamental measures to group and connect “symptoms”, “causes”, “remedies”, and “side-effects”, and guide the search for potential solutions to the problems. In terms of requirement analysis and system design, the proposed methodology can enable system designers to explore the decision space in a structured manner. In terms of evaluation, the proposed methodology is time-efficient and complementary to various forms of empirical studies, such as user surveys, controlled experiments, observational studies, focus group discussions, and so on. In general, it reduces the amount of trial-and-error in the lifecycle of VA system development.

%% file: Sections/4.CaseStudies.tex
\begin{figure*}[th]
    \centering
    \includegraphics[width=\linewidth]{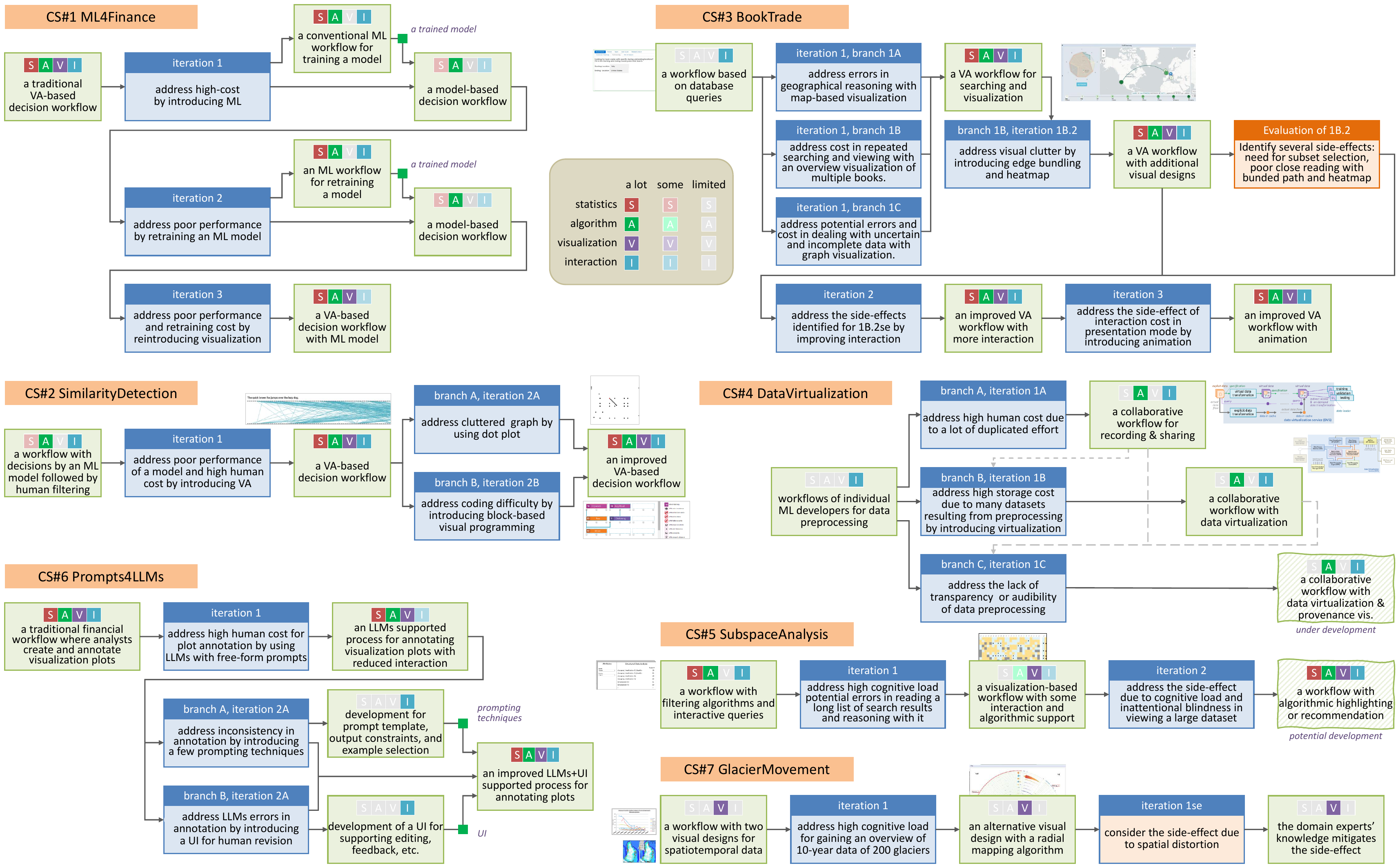}
    \caption{The workflows for optimizing VA workflows in seven case studies.}
    \label{fig:CaseStudyWFs}
\end{figure*}

\section{Actions: Seven Case Studies}
\label{sec:CaseStudies}
\subsection{A Brief Summary of Case Studies}
\label{sec:CS-Summary}
Six participants conducted seven case studies as the \emph{action} part of the research. They have varying levels of visualization expertise and different experience with VIS applications, including digital humanities, finance, glaciology, seismology, and cybersecurity. They selected the topics of case studies based on their own expertise and experience. These seven case studies are:
\begin{itemize}
    \item \textbf{CS\#1 ML4Finance} -- A participant acted as a reviewer and retrospectively examined the evolution of a financial data-to-recommendation workflow reported in two papers \cite{Yan2025VisualAnalysis,Yan2025FundSelector:Selection}.
    (See also Appendix A.)
    \item \textbf{CS\#2 SimilarityDetection} -- One participant retrospectively examined one of her past projects. The project was on detecting similarity in 18th-century French literature using VA \cite{abdulrahman:2017:CGF}.
    (See also Appendix B.)
    \item \textbf{CS\#3 BookTrade} -- One participant retrospectively examined her PhD project that was completed recently. The project was on visualizing historical records about the provenance and circulation of early printed books \cite{xing:2024:TVCG}.
    (See also Appendix C.)
    \item \textbf{CS\#4 Data Virtualization} One participant analyzed an ongoing project led by himself for developing a VIS4ML infrastructure. The study was partly retrospective on the completed parts of project \cite{Khan:2025:SCC} and partly prospective on the planned development.
    (See also Appendix D.)
    \item \textbf{CS\#5 SubspaceAnalysis} -- One participant retrospectively examined one of his PhD projects, where he developed an interactive visualization solution for finding clusters in different subspaces within a dataset \cite{jentner2023}.
    (See also Appendix E.)
    \item \textbf{CS\#6 Prompts4LLMs} -- One participant 
    acted as a reviewer and retrospectively examined a technical solution in finance for utilizing large language models (LLMs) to annotate visualization plots \cite{Hao2025}.
    (See also Appendix F.)
    \item \textbf{CS\#7 GlacierMovement} -- One participant retrospectively examined one of her projects. In the project, a new visual design was developed to provide overview visualization for spatiotemporal data in glaciology \cite{drocourt-2011}.
    (See also Appendix G.)
\end{itemize}

After familiarizing themselves with the SCORE methodology, participants analyzed the development cycles in their selected case studies. In \emph{retrospective studies}, they identified the main problems that needed to be addressed and the solutions that were developed. They considered each development cycle as an iteration of improving an existing workflow. They mapped problems to \emph{symptoms} and solutions to \emph{remedies}, followed by considering \emph{causes} and \emph{side-effects}.

The participants in two studies conducted a prospective analysis of future development. In \textbf{CS\#4 Data Virtualization}, the participant translated the not-yet-implemented requirements in a project to symptoms, and continued to reason about causes, remedies, and side-effects. In \textbf{CS\#5 SubspaceAnalysis}, the participant related the limitations reported in his paper \cite{jentner2023} as side-effects, and by treating the side-effects as new symptoms, he continued to reason about causes, remedies and side-effects for future development. 

Fig. \ref{fig:CaseStudyWFs} depicts the development cycles for each case study. In the figure, each light green box represents an existing workflow with problems or a new workflow with solutions, and each blue box represents a development cycle, which we will refer to as a \textbf{workflow} for optimizing VA workflows. Note that in some detailed case study reports in Appendices A--G, some workflows in Fig. \ref{fig:CaseStudyWFs} are further decomposed into sub-workflows. 

All participants made a serious effort to use the terms in the SCORE methodology, such as \emph{alphabet compression}, \emph{potential distortion}, and \emph{costs}. Most managed to translate the concrete entities of symptoms, causes, remedies, and side-effects into abstract entities in the form of [stat, alg, vis, int] $\times$ [high, low] $\times$ [AC, PD, Ct] (e.g., int\_high\_Ct, alg\_high\_PD, vis\_low\_AC).

\subsection{Observations}
Through seven case studies, we can make several observations.

\vspace{2mm}\noindent
\textbf{O$_1$: SCORE has been applied to VA workflows for a diversity of application domains.}
The seven case studies cover a range of applications, including:
\begin{itemize}
    \item Finance -- \textbf{CS\#1 ML4Finance}, \textbf{CS\#6 Prompts4LLMs},
    \item Humanities -- \textbf{CS\#2 SimilarityDetection}, \textbf{CS\#3 BookTrade},
    \item Data Infrastructure -- \textbf{CS\#4 DataVirtualization},
    \item Data Mining -- \textbf{CS\#5 SubspaceAnalysis}, and
    \item Geoscience -- \textbf{CS\#7 GlacierMovement}.
\end{itemize}

In addition, the application domains of previous case studies on the IVAS website \cite{IVAS:2026:web} also include: pandemic modeling \cite{Rydow:2023:TVCG}, law enforcement \cite{Zhao:2017:THS}, social media data analysis \cite{Zhang:2016:CGF}, music \cite{Ye:2023:TVCG}, generic visual designs \cite{Zhang:2018:CHI,Jin:2024:TVCG} and dashboard design \cite{Bach:2023:TVCG}.

 % ML infrastructure (CS4 – DataVirtualization: machine learning support), exploration of structured data (CS5 – SubspaceAnalysis: multi-dimensional pattern exploration), financial narrative visualization (CS6 – Prompts4LLMs: large language model-supported graphical overlays), and geoscience (CS7 – GlacierMovement: spatio-temporal glacier dynamics). The set includes workflows with a predominantly algorithmic bottleneck (ML4Finance, DataVirtualization, SubspaceAnalysis), a visual bottleneck (BookTrade, GlacierMovement), or a mixed bottleneck (SimilarityDetection, Prompts4LLMs). Notably, DataVirtualization focuses on an infrastructure workflow in which virtualization is utilized only as a late-stage remedy for lost provenance, whereas ML4Finance traces the interaction between ML model development and deployment, thereby demonstrating applicability beyond classical design-study contexts.

\vspace{2mm}\noindent
\textbf{O$_2$: There were different workflow modes: Iteration, Nesting, and Branching.}
The original SCORE methodology only presented the iteration approach, where a new iteration was driven by the discovery of side-effects in the previous iteration. During our discussions of case studies, the team started to question if there were other approaches after \textbf{CS\#4 DataVirtualization} provided a counter-example, where the three branches of workflows were planned to be developed in parallel from the beginning (Fig. \ref{fig:CaseStudyWFs}[middle-right]). Clearly, the remedies were not discovered iteratively.
The team realized that the iteration-only mindset was likely conditioned by the common scenario in which one VIS researcher executed multiple development cycles in sequence, giving an impression of iteration.

Once the team determined the need to introduce the branching approach, we also identified such an approach in \textbf{CS\#2 SimilarityDetection} (Fig. \ref{fig:CaseStudyWFs}[middle-left]) and \textbf{CS\#3 BookTrade} (Fig. \ref{fig:CaseStudyWFs}[top-right]), which were previously drawn as iterations. Subsequently, the participant of \textbf{CS\#6 Prompts4LLMs} identified a branching approach in the case study without much confusion.

Meanwhile, iterations driven by side-effects do exist, exemplified by
\textbf{CS\#1 ML4Finance} (Fig. \ref{fig:CaseStudyWFs}[top-left]),
\textbf{CS\#5 SubspaceAnalysis} (Fig. \ref{fig:CaseStudyWFs}[3rd row, right]), and
\textbf{CS\#7 GlacierMovement} (Fig. \ref{fig:CaseStudyWFs}[bottom-right]).
Furthermore, some case studies feature both iteration and branching approaches, such as
\textbf{CS\#2 SimilarityDetection} (Fig. \ref{fig:CaseStudyWFs}[middle-left]) and \textbf{CS\#3 BookTrade} (Fig. \ref{fig:CaseStudyWFs}[top-right]), and
\textbf{CS\#6 Prompts4LLMs} (Fig. \ref{fig:CaseStudyWFs}[bottom-left]).

The team also considered Munzner's nested model \cite{Munzner:2009:TVCG} and concluded that this differed from iteration and branching. Stimulated by case studies (i.e., actions), the discussions led to the formulation of a generic workflow to optimize VA workflows (Section \ref{sec:W4OVAW}).

\vspace{2mm}\noindent
\textbf{O$_3$: SCORE can be used by VIS researchers with different levels of expertise or experience.}
The participants came into this action research project with a plethora of complementary expertise (e.g., VIS, information theory, grounded theory, data mining, machine learning, visualization psychology, evaluation methods, software engineering, infrastructure development, etc.) and diverse experience of different applications (e.g., digital humanities, finance, glaciology, semiology, cybersecurity, etc.). Although participants have different levels of involvement in VIS, ranging from 2 to 30+ years, the team found that the fluency in applying SCORE to a practical problem and reasoning with abstract entities did not depend on their VIS experience in general.

Meanwhile, some participants found the step of \emph{instantiation} more demanding. It often took more effort to ``imagine'' different concrete entities related to an abstract entity. We conjectured that this capability might be expertise- and experience dependent. For example, a person with more expertise in visual design but less expertise in data mining may find it easier to instantiate abstract entities related to ``Vis'' (visualization) than ``alg'' (algorithms). This led to the conclusion that ideally, a software tool could store a database of concrete entities and prompt VIS researchers and practitioners to provide possible concrete entities when given an abstract entity.   
%
% The participant undertaking ML4Finance and Prompts4LLMs specializes in information theory, finance, and machine learning workflows. SimilarityDetection is conducted by an expert on information visualization, Human-computer Interaction, and visualization in digital humanities. Furthermore, BookTrade is carried out by a researcher with qualifications in computer science, data science, statistics, and applied psychology; the senior scientist conducting DataVirtualization focuses on the intersection of machine learning, software engineering, and data visualization; the visual analytics expert for SubspaceAnalysis researches ML and AI integration with visual interfaces; and GlacierMovement is conducted by a senior researcher with expertise in data visualization, visual analytics, human factors, and high-performance computing. The diverse backgrounds, experiences, and qualifications of the domain experts ensure a cross-disciplinary assessment informed by different design traditions.

\vspace{2mm}\noindent
\textbf{O$_4$: SCORE can be used for \emph{retrospective} and \emph{prospective} analysis.}
All seven case studies featured retrospective analysis, i.e., SCORE was applied to VA workflows that had already been completed and reported in published papers. Although most of the workflows involved were developed by participants themselves, in two case studies, (i.e., \textbf{CS\#1 ML4Finance} and \textbf{CS\#6 Prompts4LLMs}, the participant acted as an independent reviewer of previously-unknown works. This was similar to the three case studies reported in Chen and Ebert's original paper \cite{Chen:2019:CGF}, where an author acted as an independent reviewer of previously-unknown works. Our case studies further supported the conjecture that SCORE can be used by authors to reason their works systematically and for reviewers to evaluate such reasoning independently.
Studies involving retrospective analysis illustrated the merits of SCORE in clarifying design rationales and unaddressed side effects.

Two case studies (\textbf{CS\#4 DataVirtualization}, \textbf{CS\#5 SubspaceAnalysis}) featured elements of prospective analysis, i.e., participants used SCORE to explore possible remedies and select a solution before any implementation. Together with some previously reported case studies (e.g., \cite{Ye:2023:TVCG,Jin:2024:TVCG}), they suggested the potential of deploying SCORE at different stages of a VA project, including requirement analysis (i.e., analyzing symptoms and causes), solution design (i.e., exploring remedies, anticipating and addressing side-effects), evaluation (e.g., analyzing user feedback), and improvement (i.e., applying SCORE again). Studies involving prospective analysis showed that SCORE can guide forward-looking remedies and design decisions.

% Five case studies (ML4Finance, SimilarityDetection, BookTrade, Prompts4LLMs, GlacierMovement) are conducted retrospectively: the method is applied to published systems to reconstruct and, in part implicitly, improve the symptom–cause–remedy–side-effect chains. While two case studies have both prospectively and retrospectively performed parts (DataVirtualization, SubspaceAnalysis). In DataVirtualization, branches A and B have been performed retrospectively, while the provenance-visualization component (branch C), still under development, constitutes the prospective analysis. Subspace Analysis is predominantly conducted retrospectively, although it includes a prospectively analyzed component for potential development, with algorithmic highlighting or recommendations. 

% Studies involving retrospective analysis illustrate the method’s applicability in clarifying design rationales and unaddressed side effects. The prospective analysis shows that the same reasoning can also guide forward-looking remedies and design decisions.

\vspace{2mm}\noindent
\textbf{O$_5$: SCORE can derive VIS as well as non-VIS remedies.}
As shown in Fig. \ref{fig:CaseStudyWFs}, when excluding all initial benchmark workflows in the seven case studies, most improved workflows include VIS as remedies. This observation is likely biased by the strong VIS experience of the project team and should not be generalized.

Nevertheless, there are interesting exceptions. In \textbf{CS\#4 DataVirtualization}, the first two branches with known solutions do not involve VIS as remedies. In \textbf{CS\#5 SubspaceAnalysis}, the remedy suggested by the prospective analysis is primarily an algorithmic solution. This showed that SCORE itself is not biased towards VIS solutions and can be used to improve workflows with VIS and non-VIS remedies.

Another significant case study is \textbf{CS\#1 ML4Finance}, where the baseline workflow is a VA workflow. The first two iterations resulted in workflows with limited interactive visualization, and the third iteration significantly improved visualization in the workflow. The case study suggests that if SCORE were used in the first or the second iteration, the developers could identify the side-effects of the machine-centric workflows earlier and possibly derive an improved VA workflow with ML models without the delay in arriving at the final workflow after implementing two less-satisfactory workflows.%

% The case studies differ significantly in the proportion of remedies that use visualization. BookTrade and GlacierMovement are primarily visual, with both introducing new representations (geo-based provenance maps and radial spatio-temporal mappings) and algorithmic remedies in supporting roles. ML4Finance represents a trajectory in which an algorithmic remedy (ML-based ranking) is later supported by coordinated visualizations. SimilarityDetection advances a machine-centric baseline to a VA workflow with interactional and visual remedies. At the opposite extreme, DataVirtualization operates on a baseline process in which visualization is absent, utilizing mainly algorithmic remedies, and introducing visualization only as a final remedy to facilitate provenance transparency. Prompts4LLMs integrates visual, statistical, algorithmic, and interactional remedies across its branches. These findings demonstrate the method’s generalizability, supporting workflows across the entire machine-human spectrum.

\vspace{2mm}\noindent
\textbf{O$_6$: The trade-offs between alphabet compression, potential distortion, and cost occur in all case studies.}
This cross-sectional observation deserves attention, supporting the original theoretical postulation that these are three fundamental abstract measures \cite{Chen:2016:TVCG}. For example, in \textbf{CS\#1 ML4Finance}, a remedy with an ML model (algorithm and statistic) was used to address the cost of visualization and interaction. However, due to excessive alphabet compression by the ML model, the potential distortion increases. Finally, visualization was significantly improved to address potential distortion, while the ML model was used to reduce the cost of visualization and interaction. For detailed discussions on such a trade-off, see individual case study reports in Appendices A $\sim$ G.

%% file: Sections/5.HypotheseAnalysis.tex
\section{Hypotheses Evaluation}
\label{sec:HypothesesEvaluation}
While conducting their case studies, the six participants formulated their individual reflections on the five hypotheses outlined in Section \ref{sec:Methods}. They shared their reflections with the team as part of the presentation of case studies. The team discussed all five hypotheses and their evaluation during several online meetings. In this section, we report our consensus, notable disagreements, and collective insights on these hypotheses. 

% The five established hypotheses (H1-H5) of this research study were investigated through a series of collaborative meetings in which domain experts presented a prepared case study to the group. Following this presentation, the experts responded to each hypothesis and conducted a structured discussion on the implications arising from their experiences. After completing all case study sessions, the participants’ individual reflections on the hypotheses were synthesised and are reported in this section. The experience of the domain experts spans VA system design, visual analytics, data science, machine learning, digital humanities visualisation, software engineering and information theory. Instead of reporting each expert’s response in isolation, we report arising consensus, notable disagreements, and joint insights that inform the method improvements outlined in this paper. 

\subsection{Knowledge Prerequisites (Hypotheses 1 and 2)}

The first hypothesis is concerned with whether a practitioner needs to know information theory and, if so, how much. Five of the six participants advised that basic familiarity with information-theoretic concepts is required. All five agreed that a qualitative understanding can be sufficient. From their own actions, participants inferred that VIS researchers and practitioners can perform reasoning about \emph{Alphabet Compression}, \emph{Potential Distortion}, and \emph{Costs} at a conceptual level.
% and analyzing the implications of one measure propagating through subsequent components in workflows.
They do not necessarily require an advanced formal mathematical understanding of Shannon entropy or information-theoretical divergence measures.
One participant presented a different assessment, arguing that reasoning about \emph{Symptoms}, \emph{Causes}, \emph{Remedies}, and \emph{Side-effects} can be applied as a structured design vocabulary independent of the information-theoretic measures and abstraction.
Another noted an initial uncertainty about the information-theoretical concepts, suggesting that they could pose a barrier even when the fundamental principles are attainable through exemplifications. The participant suggested that some practical examples and intuitive metaphors that exemplify information-theoretic concepts could be beneficial to teach the three measures. Although case studies conducted in this work and previously can provide practical examples, we will address the metaphor need in Section \ref{sec:Discussions}.

The second hypothesis is concerned with the extent to which psychology knowledge is needed. All six participants agreed that a fundamental understanding of human cognition is valuable, specifically some knowledge of cognitive load, working memory capacity, perceptual biases, and perception of visual representations. Such knowledge is most relevant when reasoning about potential distortion and costs in human-centric processes (e.g., visualization and interaction), but also in terms of understanding and trust in dealing with statistical measures and algorithmic decisions in machine-centric processes. From their own experience, several participants inferred that VIS researchers and practitioners would be able to estimate cognitive costs based on their individual knowledge and experience, although the accuracy would depend on the level of such knowledge and experience. One participant emphasized that the absence of such conceptual familiarity would not impede the use of SCORE but would result in a prolonged SCORE process due to the need to seek external advice on human cognition. The consensus of the team is that VIS researchers and practitioners do not require formal training in psychology, but can benefit from functional knowledge of aspects of applied psychology and cognitive phenomena related to human performance in VA processes.

\subsection{Method Description and Clarity (Hypothesis 3)}

All six participants suggested that SCORE could be better described, although their suggestions for improvements varied significantly. One recurring finding in several case studies is that the notion of iteration needs to be clarified, since there is confusion about whether branching and nesting are iteration or not. Most of the participants expressed a desire to have some step-by-step guidance during the familiarization stage of using SCORE. Some participants called for a concise procedural summary. Following the suggestions, the team formulated a generic workflow to optimize VA workflows (Section \ref{sec:W4OVAW}).

Another recurring finding in different case studies is the need to make the instantiation easier. Several participants suggested that one introduce a ``mapping dictionary'' or a large repository of examples that would help VIS researchers and practitioners to find appropriate practical instances (concrete entities) when they encounter any of the 24 abstract entities. All team members consider this need a critical part of our roadmap for future development (Section \ref{sec:Discussions}).

However, one participant provides a contrasting view, arguing that over-formalizing the SCORE methodology would be self-defeating, and instead, SCORE should be interpreted as a flexible vocabulary for communicating design trade-offs rather than a prescriptive, fixed protocol. The sweet spot between structured guidance and methodological flexibility implies that SCORE description may offer scaffolding, a combination of existing templates, worked examples, and potential checklists without the strict constraints imposed by a fixed sequence.
%In Section \ref{sec:Discussions}, we respond directly to this finding by offering metaphor analysis and further guidance.

% Some participants called for a concise procedural summary that guides practitioners through workflow segments, abstracting and evaluating symptoms, utilising the adjacency matrices, and predicting side effects \cite{Chen:2019:CGF}.

% The sessions also emphasised the value of metaphors again, with one expert noting that the “close versus distant reading” contrast from digital humanities offers an effective way to investigate the trade-off between close inspection and high-level abstraction. 

% However, another participant provides a contrasting view, arguing that over-formalising the method would be self-defeating and should instead be interpreted as a flexible vocabulary for communicating design trade-offs rather than a prescriptive, fixed protocol. The mutual concession between structured guidance and methodological flexibility implies that the method description should offer scaffolding, a combination of existing templates, worked examples, and potential checklists without the strict constraints of imposing a fixed sequence. In Section \ref{sec:Discussions}, we respond directly to this finding by offering metaphor analysis and further guidance.

\subsection{Procedural Extensions (Hypothesis 4)}

All six participants affirmed the need for additional procedures. During the discussions, one proposed a systematic retrospective review procedure to verify that critical symptoms were investigated, serving as a completeness check. Another suggested relating SCORE to the traditional requirements-evaluation workflows, so that inexperienced users can relate information-theoretic cost-benefit reasoning to the traditional design study practices. Several participants underscored the need for practical guidance in applying SCORE under real-world constraints, including incomplete data or multiple stakeholders with divergent priorities, which in theory had been covered by information theory and cost-benefit analysis \cite{Chen:2016:TVCG}.

Furthermore, there have not been many case studies on disseminative visualization largely due to the previous focus of SCORE on VA. In \textbf{CS\#3 BookTrade}, one user requirement was for engaging the public. This suggests further research into the application of information theory and SCORE to disseminative visualization, addressing topics such as aesthetics, attention, and trust.

% The domain experts reasoned that the method requires additional case studies from diverse contexts. These observations motivate the case studies summarised in Section \ref{sec:CaseStudies}, iterative, nested, branching, and comparison workflow configurations and a resulting workflow for optimising VA workflows (WF4OWF) approach outlined in Section \ref{sec:W4OVAW}.

\subsection{Other Means for Improvement (Hypothesis 5)}
All participants agreed that it would be useful to provide SCORE with software assistance. In addition to aforementioned ``mapping dictionary'' and a large repository of examples, the team members also suggested (i) a user interface for navigating the 24 abstract entities and their relations interactively with graphs or adjacency matrices; (ii) prompts and recommendations for guiding SCORE users; (iii) an evidence and assumption log, (iv) provenance visualization for previously reasoned workflows and their symptoms, causes, remedies, and side-effects; and (v) some assessment mechanisms, e.g., multi-criteria decision analysis, for comparing different remedies. Some team members also speculated that when there are a large number of case studies in the future, AI can learn to perform reasoning and make recommendations, providing assistance to VIS researchers and practitioners. All of these are featured on our proposed roadmap as described in Section \ref{sec:Discussions}.

%% file: Sections/6.W4OVAW.tex
\begin{figure*}[th]
    \centering
    \includegraphics[width=168mm]{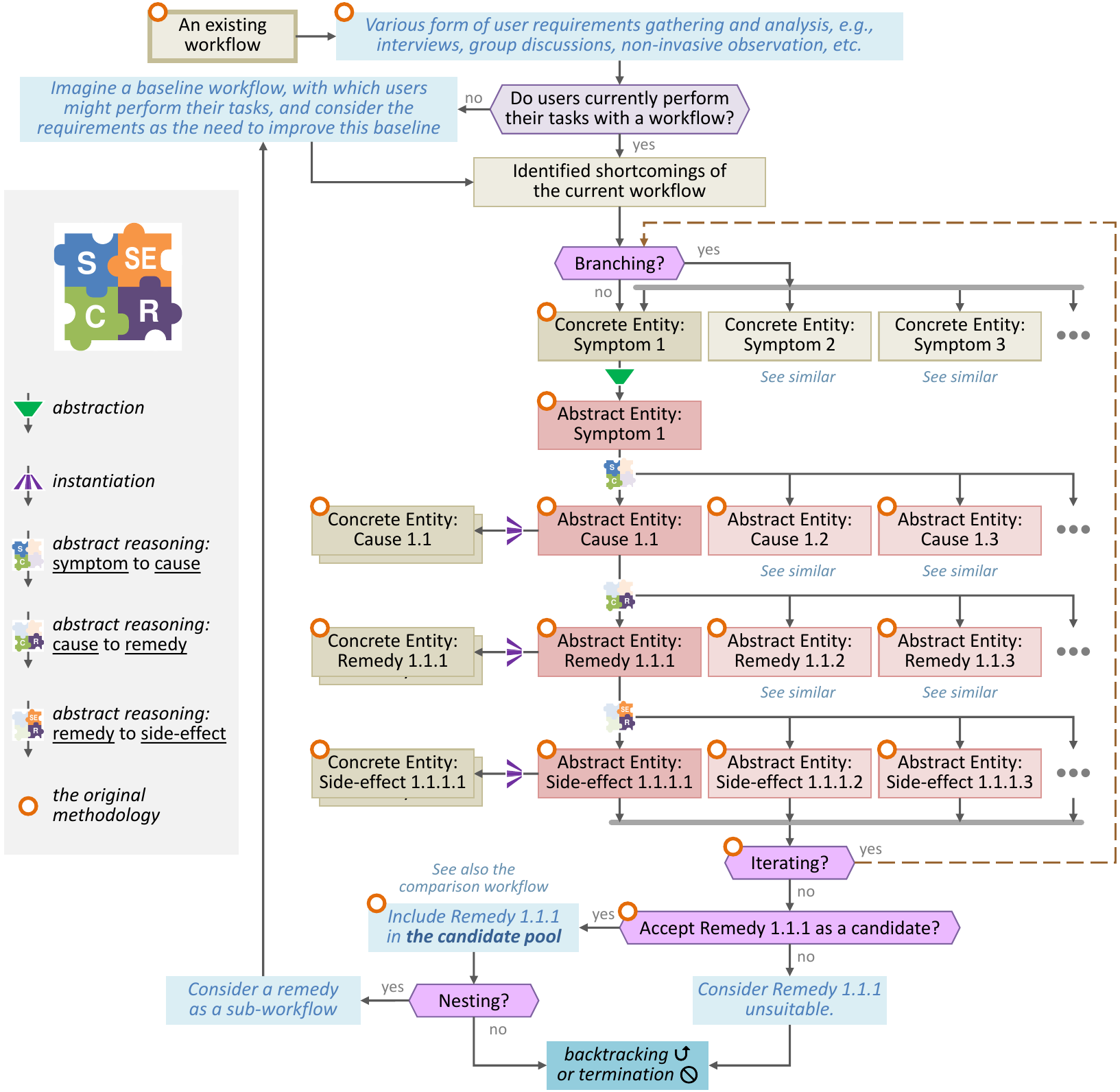}
    \caption{The main workflow for optimizing VA workflows}
    \label{fig:W4OVAW}
\end{figure*}

\section{Workflow for Optimizing VA Workflows}
\label{sec:W4OVAW}

As summarized in Section \ref{sec:HypothesesEvaluation}, the participants identified that it would be helpful to accompany the methodology with some form of step-by-step guidance. Furthermore, the analysis of the case studies indicated the need to consider different purposes and approaches within the broad term ``iteration'', such as branching, decomposition, and comparison. These led to the formulation of a generic workflow for optimizing VA workflows.%

\begin{figure}[th]
    \centering
    \includegraphics[width=70mm]{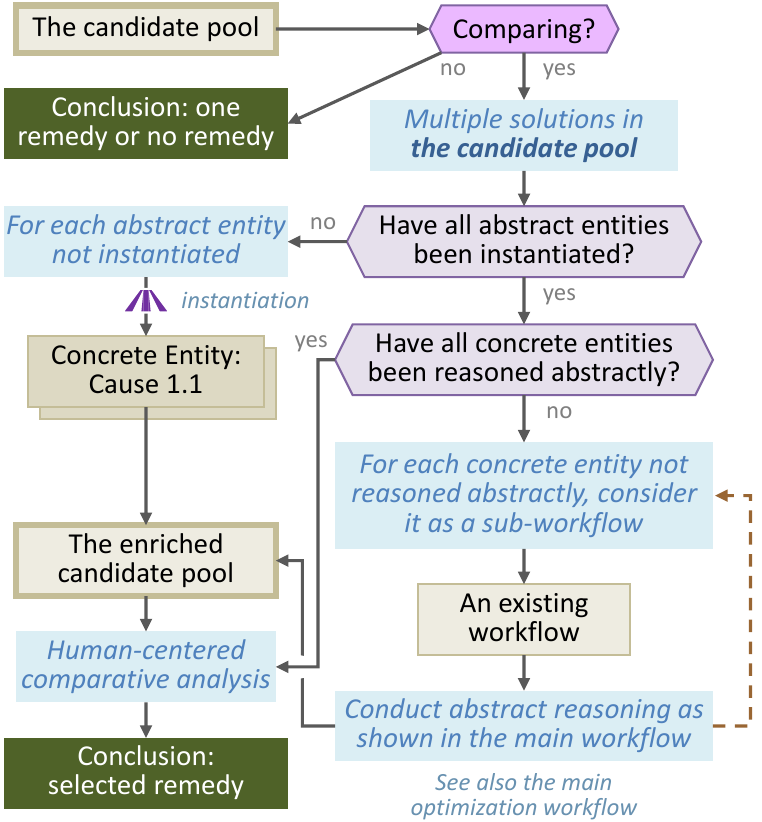}
    \caption{The additional comparison workflow for optimizing VA workflows}
    \label{fig:Comparison}
\end{figure}

As shown in Fig. \ref{fig:W4OVAW}, the generic workflow includes the major components of the original SCORE methodology (indicated by orange circles), although the original paper \cite{Chen:2019:CGF} does not present these components as a workflow. This generic workflow includes new components related to branching and nesting, as well as defining a baseline workflow when it was not given. In addition, Fig. \ref{fig:Comparison} provides a sub-workflow for comparing different remedies. While this generic workflow provides step-by-step guidance, it is not meant to be executed rigidly and should indeed be considered as a step-by-step recommendation.

\vspace{2mm}\noindent
\textbf{Iteration for Decomposition.} A VA workflow usually consists of different processes that can be machine-centric (e.g., statistics and algorithms) and human-centric (e.g., visualization and interaction). Following a high-level optimization, one may continue to optimize a specific process by considering it as a sub-workflow to be optimized. For example, in the case study \textbf{CS\#5 SubspaceAnalysis}, after a VIS process was introduced to the otherwise machine-centric workflow, one continued to optimize the VIS process as a sub-workflow by introducing the use of a sorting algorithm as discussed in Appendix E. 

In the context of visualization, the nested model focuses on iterations based on decomposition (e.g., encoding/interaction technique design followed by algorithm design) \cite{Munzner:2009:TVCG}. Here, the main differences are: (i) both machine- and human-centric workflow and sub-workflows can be further decomposed.
(ii) there is neither restriction in terms of the number of levels of decomposition nor semantic constraint as to the functionality of each level.

\vspace{2mm}\noindent
\textbf{Iteration for Side-effects.} This type of iteration is the main focus of the original methodology, which was briefly mentioned as refinement in the nested model \cite{Munzner:2009:TVCG}. The case study \textbf{CS\#1 ML4Finance} represents this type with three major iterations inspired by two papers \cite{Yan2025VisualAnalysis,Yan2025FundSelector:Selection}.

\vspace{2mm}\noindent
\textbf{Branching.} For a workflow with several shortcomings, one may address different shortcomings in parallel. In \textbf{CS\#4 DataVirtualization}, three major technical solutions were identified in the requirements analysis and design specification stages, and parallel implementations were planned.

When such implementations are carried out by very limited development resources (e.g., one developer), different branches are usually done sequentially, giving an impression of iteration. Indeed, \textbf{CS\#3 BookTrade}, which features a complex development life cycle \cite{xing:2024:TVCG}, was initially perceived to have many iterations of side-effects. The presentation of \textbf{CS\#4 DataVirtualization} and the following team discussions led to the discovery of multiple types of optimization progression. 

\vspace{2mm}\noindent
\textbf{Comparison.} When there are two or more candidate remedies, it is necessary to conduct a comparative analysis with both abstract and concrete entities. If candidate remedies were identified without using abstract reasoning (e.g., the sub-workflow in \textbf{CS\#7 GlacierMovement}), one can carry out abstract reasoning to identify potential side-effects and potential solutions to address such side-effects systematically before the comparison. If candidate remedies were identified using abstract reasoning, their corresponding concrete entities identified by instantiation must be available in the comparative analysis.

This principle can also be explained by information theory. Abstraction is a many-to-one mapping, and comparison with abstract entities will likely lead to potential distortion. Instantiation provides a more precise reverse mapping, allowing humans to bring their knowledge about the specific concrete entities into the comparative analysis, hence reducing potential distortion.

%% file: Sections/7.Discussions.tex
\section{Discussions}
\label{sec:Discussions}
Our actions (case studies) as reported in Section \ref{sec:CaseStudies} and action analysis (hypotheses evaluation) as reported in Section \ref{sec:HypothesesEvaluation} led us to the formulation of the generic workflow as described in the previous section. In this section, we focus on two potential measures that will facilitate the broad application of the SCORE methodology as suggested by action analysis. The first measure is to identify intuitive metaphors that can be used to assist in learning information-theoretic concepts that underpin the SCORE methodology. The second measure is to establish a roadmap for developing software that helps VIS researchers and practitioners to use SCORE in practice.  

% The results of our hypotheses evaluation and the case studies suggest two complementary directions for enhancing the SCORE method: conceptual advancement to enable broader accessibility, and technological development to reduce application costs in practice. To clarify, this section first introduces metaphors that emerged during our discussion meetings, which serve as tools for conceptualising the information-theoretic frameworks supporting the method, and then presents the findings of a survey investigating the efficacy of these metaphors. Finally, we outline a road map for developing software tools to assist practitioners in analysing, modelling, and recording the optimisation of VA workflows. 

\begin{table*}[ht!]
\centering
\caption{%
  Metaphor Survey Results with average ratings (with min and max in parentheses) for ten metaphors. 
  M1: Close Reading vs.\ Distant Reading;
  M2: Commercial Case Study vs.\ Market Analysis;
  M3: Consumer Interview vs.\ Market Segmentation;
  M4: Topographic Survey (One Area) vs.\ Global Model (Continental-Scale);
  M5: Street-Level View vs.\ Map View;
  M6: Fieldwork vs.\ GIS Analysis;
  M7: Microscope vs.\ Satellite Imaging;
  M8: Time-Series Zoom-In vs.\ Zoom-Out;
  M9: Clinical Case Study vs.\ Epidemiological Study;
  M10: Four Professors' Debate on Processing Exam Results (Stat.\ vs.\ Alg.\ vs.\ Vis.\ vs.\ Interaction).%
}
\label{table:metaphor}
\renewcommand{\arraystretch}{1.6}
\setlength{\tabcolsep}{5pt}
{\setlength{\arrayrulewidth}{0.3pt}%
\scalebox{0.78}{%
\begin{tabular}{|l|c|c|c|c|c|c|c|c||c|c|c|c|}
 \hline
  & \textbf{AC} & \textbf{PD} & \textbf{Cost} & \textbf{Knowledge} & \textbf{Statistics} & \textbf{Algorithms} & \textbf{Visualization} & \textbf{Interaction}
  & \textbf{Most ${>}75\%$} & \textbf{Many ${>}50\%$} & \textbf{Some ${>}25\%$} & \textbf{Few ${\leq}25\%$} \\
 \hline
 M1  & 4.50 (3, 5) & 4.25 (1, 5) & 4.38 (4, 5) & 4.00 (1, 5) & 3.50 (1, 5) & 3.50 (3, 4) & 4.13 (3, 5) & 3.25 (2, 4) & 4 & 3 & 1 & 0 \\
 \hline
 M2  & 4.00 (1, 5) & 4.00 (2, 5) & 4.13 (2, 5) & 3.50 (1, 5) & 4.00 (2, 5) & 3.13 (2, 4) & 3.75 (1, 5) & 3.13 (1, 5) & 2 & 4 & 2 & 0 \\
 \hline
 M3  & 3.75 (2, 5) & 3.63 (1, 5) & 3.88 (2, 5) & 2.88 (1, 5) & 3.38 (1, 5) & 3.38 (2, 5) & 3.50 (1, 5) & 2.63 (1, 5) & 2 & 2 & 4 & 0 \\
 \hline
 M4  & 4.00 (2, 5) & 4.13 (2, 5) & 4.25 (3, 5) & 3.50 (2, 5) & 3.75 (2, 5) & 3.50 (2, 5) & 3.88 (2, 5) & 3.38 (2, 5) & 1 & 5 & 1 & 1 \\
 \hline
 M5  & 4.13 (1, 5) & 3.88 (1, 5) & 4.00 (3, 5) & 3.38 (1, 5) & 2.88 (1, 4) & 3.00 (2, 4) & 3.88 (1, 5) & 3.38 (1, 5) & 7 & 1 & 0 & 0 \\
 \hline
 M6  & 3.75 (2, 5) & 2.88 (1, 5) & 4.00 (1, 5) & 3.13 (1, 5) & 3.63 (2, 5) & 3.38 (1, 5) & 3.50 (1, 5) & 3.13 (1, 4) & 0 & 4 & 4 & 0 \\
 \hline
 M7  & 4.13 (2, 5) & 3.38 (1, 5) & 3.63 (2, 5) & 3.00 (2, 4) & 2.38 (1, 4) & 2.63 (1, 4) & 3.63 (1, 5) & 3.00 (1, 5) & 5 & 2 & 1 & 0 \\
 \hline
 M8  & 4.63 (2, 5) & 3.75 (1, 5) & 3.88 (1, 5) & 3.38 (1, 5) & 3.38 (1, 5) & 3.00 (1, 4) & 3.88 (2, 5) & 3.63 (1, 5) & 7 & 1 & 0 & 0 \\
 \hline
 M9  & 4.13 (1, 5) & 4.00 (2, 5) & 3.25 (1, 5) & 3.50 (1, 5) & 4.00 (2, 5) & 3.25 (1, 4) & 3.38 (1, 4) & 2.75 (2, 4) & 0 & 3 & 5 & 0 \\
 \hline
 M10 & 3.00 (1, 5) & 3.13 (1, 5) & 3.75 (1, 5) & 3.50 (1, 5) & 3.63 (1, 5) & 3.63 (1, 5) & 3.63 (1, 5) & 3.88 (1, 5) & 1 & 3 & 1 & 3 \\
 \hline
\end{tabular}%
}%
}
\end{table*}

\subsection{Discussions on Metaphors}

% \textcolor{red}{1. Focus on how to connect benefits (more broadly), potential distortion, and costs with the metaphors in this section.}

% \noindent \textcolor{red}{2. Furthermore, “4 professor problem” relates specifically to the current metaphoric discussion section/how it should be included.}

% \noindent \textcolor{red}{3. (For Min's attention): The draft below is missing some additional information (such as cost, benefit, and potential distortion) as it currently only covers the discussion about alphabet compression.}

% \noindent \textbf{\textcolor{red}{4. What is the Four Professors' Debate? could include a short description}}

The SCORE methodology was underpinned by information theory, and some information-theoretic concepts can be difficult to grasp. During the presentation of a case study in digital humanities, a participant used the metaphor ``close reading vs. distance reading'' to convey the three fundamental concepts: \emph{alphabet compression}, \emph{potential distortion}, and \emph{costs}. This stimulated some discussions in the meeting and a thread of emails afterwards. Some other metaphors were proposed, e.g., ``image compression and decompression'' and ``zoom-in vs. zoom-out''. A team member asked ChatGPT (with the GPT-5.3 model) to use different metaphors similar to ``close reading vs. distance reading'' but in different disciplines. In addition, we also considered the metaphor ``the four professors' problem'' that was mentioned in several presentations on information theoretic cost-benefit analysis.

The team first analyzed these metaphors through discourses on their suitability.  

\vspace{2mm}\noindent
\textbf{Close Reading vs. Distance Reading.} Applying excessive alphabet compression may lose information too quickly or heavily, resulting in high\_AC. This is similar to \emph{distant reading}~\cite{moretti2013distant}, which is a digital humanities technique that uses computational tools and statistical methods to analyze large literary collections, rather than a close examination of individual texts. 
In distant reading, results are typically shown as a few or a series of statistical measures. From a large amount of text to a few numbers, the information is compressed very quickly or excessively, thus high AC. When one interprets these with little knowledge of the original text, the interpretation may have a high potential distortion (high\_PD).

In contrast, \emph{close reading} involves a detailed analysis of texts, which is an intensive and costly process (high\_Ct). It emphasizes a thorough and detailed analysis of a specific text or phenomenon, such as a poem~\cite{AbdulRahman:2013:CGF}. In terms of cost-benefit analysis, close reading involves little or no compression, resulting in low\_AC in terms of reading, which in abstraction is considered a form of visualization.

% The Chen and Ebert's Cost-Benefit Analysis  methodology~\cite{Chen:2019:CGF} can be difficult to understand. In our online meetings and many email exchanges, we suggested metaphors to better grasp how this methodology relates to our case studies. Throughout our discussions, we draw on our knowledge and experience to identify relevant concepts or metaphors for the case studies. This started with applying close and distant readings to the analogy of AC. For example, applying excessive compression to a situation, such as condensing information too quickly or heavily, results in high AC, similar to distant reading~\cite{moretti2013distant}. Distant reading is a digital humanities technique that uses computational tools and statistical methods to analyze large literary collections, rather than focusing on a close examination of individual texts. In contrast, close reading involves a detailed analysis of texts; in scientific terms, it is an intensive interpretive process. In distant reading, results are typically shown as a series of numbers or statistics; here, if information is compressed too quickly or excessively, it produces a high AC. In contrast, close reading emphasizes a thorough and detailed analysis of a specific text or phenomenon, such as a poem~\cite{AbdulRahman:2013:CGF}, which, in a cost-benefit analysis, might involve little or no compression, resulting in raw data or low-AC visualizations.

\vspace{2mm}\noindent
\textbf{Microscope View vs. Satellite Imagery.}
Similarly, in the physical sciences, one may compare a microscope view with satellite imagery. A microscope provides a high-resolution view of a single specimen, revealing detailed local features such as textures, anomalies, and structures, usually in a slow, careful, and interpretive manner. Similarly to close reading, there is minimal alphabet compression (low\_AC).
% To improve this, algorithms, statistical methods, or interactive tools can be used to increase compression levels.
Satellite imaging, on the other hand, offers a broad view of an entire ecosystem at lower resolution per object. Compared with a microscope view, satellite imaging is high\_AC.
However, when observing some large scale patterns, such as distributions, clusters, and trends, satellite imaging is much more cost-effective (low\_Ct) compared to having a large number of microscope views.

\vspace{2mm}\noindent
\textbf{Clinical Case Study vs. Epidemiological Study.}
In healthcare, a clinical case study, which is typically qualitative and involves a small number of participants, emphasizes narrative analysis and understanding individual circumstances, thus low\_AC. In contrast, a population-level epidemiological study examines large samples, focusing on statistical patterns, risk factors, distributions, and long-term trends, thus high\_AC.

\vspace{2mm}\noindent
\textbf{Four Professors' Problem \cite{Chen:2022:BCS}.}
Four professors sat down to discuss the exam marks of the $N$ students. Professor S argues that it is best to consider a statistical measure such as mean. Professor A argues that it is the best to use an algorithm, such as sorting. Professor V argued that it is best to visualize the marks as a bar chart. Professor I argued that it is best to interact with the computer to retrieve a mark when it is needed. The four approaches all involve information loss, but at different levels of alphabet compression; all may cause potential distortion, which may be alleviated by knowledge of the students, course, or exam concerned; and each may incur less cost for some tasks but more for other tasks.

% Additional metaphors appear in geography: the street-level view versus the map view. The street-level perspective offers detailed sensory and micro-scale information (i.e., low AC), while the map view considers zoning, connectivity, and density at a larger scale, representing high AC.

% From a psychological point of view, this analogy compares a clinical case study with an epidemiological study.

\vspace{2mm}
After collecting a list of candidate metaphors, we selected a subset of those that are relatively easy to understand. We conducted a small survey within the team since all team members did not need any training on key concepts in SCORE. A Google form was used for the survey, which was completed by all eight team members. The survey included 10 metaphors, as shown in the first column of Table~\ref{table:metaphor}. For each metaphor, we asked team members to rate on a Likert scale how closely each metaphor relates to the information-theoretic concepts and the VA components in SCORE: namely alphabet compression, potential distortion, costs, knowledge factor, statistics, algorithms (inc. ML models), visualization, and interaction, using the scale: Very Strong, Good Enough, Not Sure, Not So Good, Very Weak.

% Whether comparing distant and close reading, microscope and satellite imaging, or high and low AC, each scale addresses different questions and emphasizes different aspects of analysis. One might offer contextual meaning, while the other uncovers structural patterns among variables. From there, we explored other metaphor types that could be applied to the SCORE methodology, drawing inspiration from ChatGPT. We validated the list, generated a set of metaphors for comparison, and conducted in-depth discussions to assess each one's validity. Once we confirmed the list of metaphors comparable to Chen and Ebert's Cost-Benefit Analysis, we conducted a Google Forms survey with all eight authors of this paper to determine which metaphors were most closely related to it. The survey included 10 metaphors, as shown in the first column of Table~\ref{table:metaphor}. For each metaphor, we asked participants to rate on a Likert scale how closely each metaphor relates to information-theoretic concepts and the VA component: AC, PD, Cost, Knowledge, Statistics, Algorithms, Visualization, and Interaction, using the scale: Very Strong, Good Enough, Not Sure, Not So Good, Very Weak.

Our survey results show that in terms of conveying the meaning of alphabet compression, the ``time-series: zoom-in vs. zoom-out'' metaphor has the highest average, indicating that it is the most suitable to the team. This is followed by ``close reading vs. distant reading''.
For potential distortion, the most suitable metaphor is ``close reading vs. distant reading'' followed by ``topographic survey (one area) vs. global model (continental scale)''.
For Cost, it is ``close reading vs. distant reading'' with ``topographic survey (one area) vs. global model (continental scale)'' next.
Regarding the knowledge factor, it is again ``close reading vs. distant reading''.
For Statistics, two metaphors score the highest mean: ``commercial case study vs. market analysis'' and ``clinical case study vs. epidemiological study''.
For Algorithms, it is the ``debate among four professors on processing examination results''. For Visualization, the most suitable metaphors are ``close reading vs. distant reading''.
For Interaction, it is the ``four professors' problem''. We also asked which of these metaphors would be most easily understood as an analogy to the key concepts in the SCORE methodology by VIS researchers and practitioners. The highly-ranked answers were ``street-level view vs map view'' and ``time-series: zoom-in vs. zoom-out'', with a result of seven counts for ``Most >75\%'' and one for ``Many >50\%'' for the two metaphors. The least understood is ``the debate among the four professors about processing examination results'' with a result of one count for ``Some >25\%'' and three counts for ``Few ${\leq}25\%$''.

Because the sample size is small, the survey results can serve only as an indicative purpose. Nevertheless, the results of the survey suggest that metaphors are useful in conveying key concepts in SCORE. Some highly-ranked metaphors, such as ``close reading vs. distant reading'', can certainly be used in training materials for SCORE. 

%----------------------------------------------------------------
%\subsection{\textcolor{red}{\textbf{Discussions on Software - Revised Version - similar to previous version with same content but keeps the road map narrative - previous version is below 7.3}}}
\subsection{Discussions on Software}
This action research project identified that it is necessary to support the SCORE methodology with a number of software capabilities, including but not limited to (a) ``mapping dictionary'', (b) a large repository of examples, (c) a user interface for navigation of the 24 abstract entities and their relations, (d) algorithmic or model-driven prompts and recommendations, (e) log recording for evidence and assumption, (f) provenance visualization, (g) some assessment mechanisms, and (h) AI-based reasoning and recommendations. (See also Section \ref{sec:HypothesesEvaluation}.)

In addition, while discussing a potential software tool, team members also identified other useful software functions, including: (i) workflow templates, construction, and editing, (j) multi-user collaboration, and (k) automated report generation.

\begin{figure}[t]
    \centering
    \includegraphics[width=\linewidth]{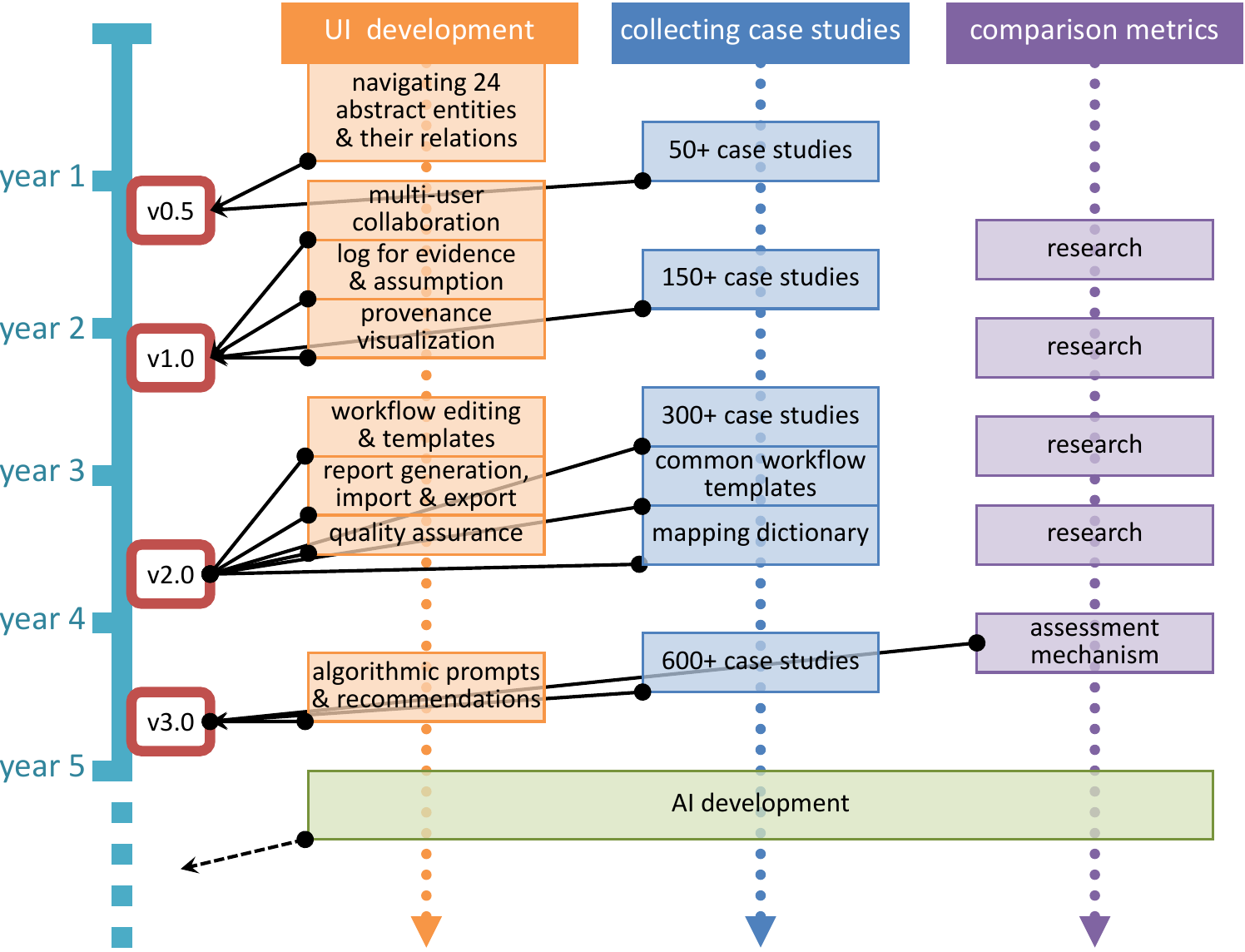}
    \caption{A roadmap for developing software to assist VIS researchers and practitioners in using the SCORE methodology.}
    \label{fig:Roadmap}
\end{figure}

Hence, a roadmap for developing such software can be outlined according to their mutual dependency and the likely development time required. Fig. \ref{fig:Roadmap} illustrates a roadmap for the development of a software gradually. The software capabilities can be roughly organized into three strands of research and development, namely \emph{UI-related development}, \emph{repository-related development}, and \emph{metrics-related development}.

One major bottleneck is expected to be the process of collecting SCORE case studies. Taking into account the seven case studies conducted in this work and several other case studies available in the literature and the IVAS website, there are fewer than 15 case studies currently. Hence, there will be a continuing effort for collecting SCORE case studies and building a \textbf{Repository} as shown along the blue strand in the middle. Building on the case studies in the repository, one can develop a number of software capabilities, including:
\begin{itemize}
    \item \emph{Workflow templates} -- one can identify common VA workflows and make them into templates to be adopted and adapted by other SCORE users.
    \item \emph{Mapping dictionary} -- one can identify common mappings from concrete entities to abstract entities and vice versa, and common transitions among the 24 abstract entities. When the size of the repository grows, the statistics of these mappings will become more reliable, facilitating the development in other strands, including algorithmic prompts and recommendations.
\end{itemize}

UI-related development includes many software capabilities that SCORE users will interact with. As shown along the orange strand in Fig. \ref{fig:Roadmap}, these capabilities include:
\begin{itemize}
    \item \emph{Navigating 24 abstract entities} -- The participants identified that working on graphs or adjacency matrices on paper demands a lot of cognitive effort and suggested that a computer-aided approach for navigating 24 abstract entities during abstract reasoning would be a large help. In addition, the UI should also support other transitions in the generic workflow to optimize VA workflows as shown in Fig. \ref{fig:W4OVAW} as well as the remedy comparison workflow as shown in Fig. \ref{fig:Comparison}.
    \item \emph{Multi-user collaboration} -- The UI supports collaborative activities involving multiple SCORE users for workflow optimization, such as working on different branches and different levels of sub-workflows, teamwork (e.g., for identifying symptoms, suggesting causes, exploring remedies, and discussing side-effects), and so on.
    \item \emph{Log recording for evidence and assumptions} -- During workflow optimization, a SCORE user will bring evidence and assumptions to justify, as well as make many proposals and decisions on symptoms, causes, remedies, and side-effects. The log will be particularly useful in multi-user collaboration, and in quality assurance.
    \item \emph{Provenance visualization} -- With multi-user collaboration and a large amount of logged data, visualizing provenance data becomes a critical capability.
    \item \emph{Workflow construction, editing, and templates} -- It is desirable to provide SCORE users with the software capability for constructing and editing VA workflow diagrams, allowing users to annotate the functionality and abstract entities of different machine- and human-centric processes in each workflow. When the repository-strand is able to provide workflow templates, they can be made available to SCORE users through the UI. 
    \item \emph{Quality assurance checklist} -- the software can also provide a mandatory quality assurance process, e.g., to verify that all abstract entities have been considered and all identified side-effects for selected remedies have been recorded.
    \item \emph{Report generation, import and export} -- To make a SCORE process auditable, SCORE users have to produce reports that can be reviewed by various stakeholders, e.g., colleagues, independent peers, and/or users of the VA workflows being improved. To ease such effort, software capabilities can be provided to generate reports automatically or semi-automatically, facilitating sharing through import and export.
    \item \emph{Algorithmic prompts and recommendations} -- When the repository-strand is able to produce relatively reliable statistics of the collected cases, it will be possible to provide SCORE users with helpful prompts and recommendations on topics such as: 
    (i) What are likely abstract entities for given concrete entities? 
    (ii) What are the likely concrete entities for given abstract entities? 
    (iii) What are typical symptoms in a common type of workflow? and 
    (iv) What are common side-effects for given remedies?  
\end{itemize}

How to compare two or more potential remedies for improving a VA workflow is a non-trivial challenge in the VIS field. Since these potential remedies have not been implemented yet, it would not be feasible to conduct any controlled user study. Interviews and group discussions may also be problematic if the target users have to ``imagine'' how different remedies may work or impact the workflow. Even if one implements only one or two remedies, any evaluation could be biased by the known remedies compared to the ``imagined remedies''. In order to address such a challenge, an extensive amount of research will be necessary. Nevertheless, some early effort has resulted in semi-quantitative methods to compare different design options in a relatively narrow context \cite{Hsieh:2025:TVCG}. We anticipate that future research will result in quantitative or semi-quantitative methods to compare potential remedies to improve VA workflows. By then, such methods can be used to enrich the SCORE methodology as well as enhance the software capability of algorithmic prompts and recommendations.

Research and development in all three strands will collectively provide a large amount of data, as well as algorithmic components to enable the development of AI solutions to optimize VA workflows.

To investigate the feasibility of the above roadmap, a few participants, who have extensive experience in designing and developing data infrastructures, ontology-based search engines, automation agents, machine learning, visualization, and user interfaces, examined the aforementioned software capabilities in some detail, including architectural layers of the system, data flows in the system, and so on. They confirmed that having an adequate case study repository would be the most critical bottleneck. Considering that there are many VA workflows reported in the literature, there is no shortage of raw materials. They hoped that the estimated number of collected case studies in Fig. \ref{fig:Roadmap} would be feasible to achieve.

%% file: Sections/8.Conclusions.tex
\section{Conclusions}
\label{sec:Conclusions}

In this paper, we adopted the Action Research method to carry out a methodological study of the SCORE methodology (Symptom, Cause, Optimize, Remedy, side-Effects) originally proposed by Chen and Ebert \cite{Chen:2019:CGF}. By conducting seven case studies as actions, participants familiarized themselves with SCORE and, therefore, were able to identify ways to improve, extend, and enrich the methodology. Before this action research project, only the original authors applied SCORE to practical problems. After the project, eight VIS researchers can now use SCORE to analyze VA workflows and identify symptoms to be addressed, determine causes, explore potential remedies for improvement, and consider side-effects associated with remedies. They also become fluent in using information-theoretic concepts to explain their analysis. With the seven new case studies, the total number of documented case studies has more or less doubled. Although the number of VIS researchers who can use SCORE and the number of SCORE case studies is still low, the project is a major step forward in transforming a predominantly theoretical framework into a practical method.

In the history of science, many theoretical ideas took time to become practical. In fact, it usually took a large amount of effort and a variety of actions by many researchers and practitioners to make theoretical ideas practical. This action research project was carried out in this spirit and resulted in many suggestions for improving the SCORE methodology. Some suggestions have been further acted upon within this project, e.g., formulating a generic workflow for optimizing VA workflows, separating the notions of branching and nesting from that of iteration, identifying suitable metaphors for conceptual components in SCORE, and recognizing the need to support SCORE with software. Some will be carried out in future work, including building a case study repository and developing a software tool.

Among many findings derived from observations and hypotheses evaluation as reported in Sections \ref{sec:CaseStudies} and \ref{sec:HypothesesEvaluation}, it is important to confirm that (i) the knowledge of SCORE and the experience of applying SCORE can be acquired, (ii) SCORE can be used to improve many VA workflows in a variety of application domains, (iii) SCORE can also be used by VIS researchers and practitioners with different levels of expertise or experience, (iv) SCORE can be used for both retrospective analysis (e.g., technical review of a piece of work) and prospective analysis (e.g., requirements analysis and design specification), and (v) the prerequisite for the knowledge of information theory and psychology is attainable. Although there are entrance barriers to becoming skilled at applying SCORE to practical problems, compared with those to becoming a medical doctor, a bridge designer, or many other professionals, these barriers are relatively easy to overcome. On the one hand, the more VIS researchers use SCORE, the more case studies they develop, the better training materials are created, and the more educators are trained. On the other hand, with more case studies, better training materials and more educators, the entrance barriers will be lower. Meanwhile, around the world, there are so many data-to-decision workflows awaiting improvement.

\newpage

%% file: Appendix/1.CaseStudyVA+ML.tex
\begin{figure*}[t]
    \centering
    \includegraphics[width=0.85\textwidth]{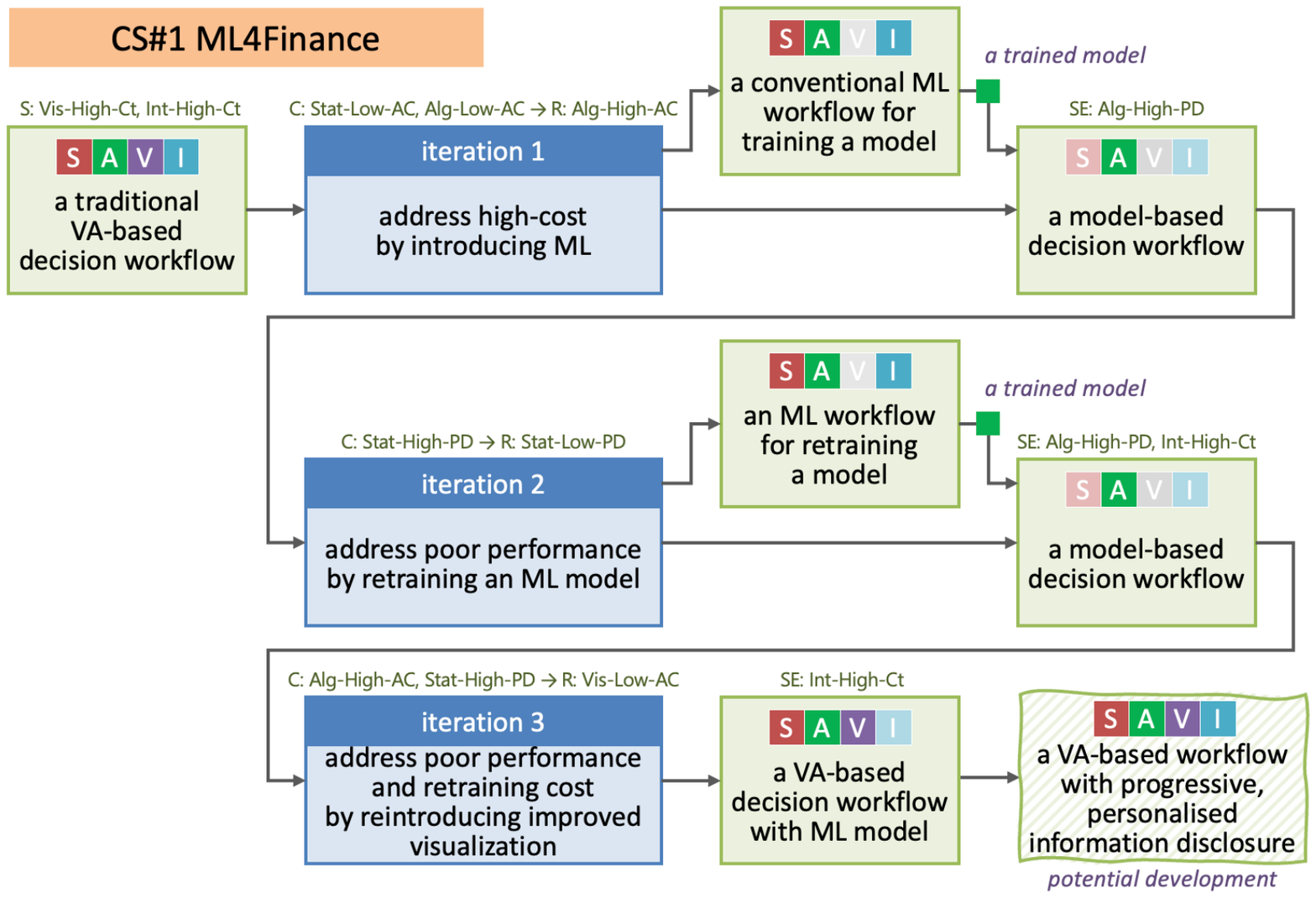}
    \caption{\textbf{CS\#1 ML4Finance}: An updated Workflow for Optimizing Workflows (WF4OWF) for the ML4Finance case. Note that this extended version includes a potential development component, explicitly marked as speculative and based solely on a co-author’s suggestion, which is not present in the main-text version in \textit{Figure} \ref{fig:CaseStudyWFs}.}
    \label{fig:cs-1}
\end{figure*}

This case study is based on the progression from Yan et al.\cite{Yan2025VisualAnalysis} to Yan et al.\cite{Yan2025FundSelector:Selection} as a post-hoc evaluation of a visual analytics design influenced by mixed-initiative machine learning. The earlier research presents the core workflow: mutual fund data are used to compute seven technical indicators covering profitability, risk resistance, and price performance, followed by training a pairwise preference-based classifier on investors' annotations of misordered funds. This workflow already combines a user interaction module with a basic visual prototype that displays fund rankings and allows adjustment of investment preferences. The latter research extends this workflow to FundSelector, a visual analytics system that combines the ranking model with six coordinated views to specify investment preferences, market context, fund rankings, indicator overviews, manager views, and adjustable temporal fund comparisons. Thus, the latter work solves multiple \textbf{symptoms} partially addressed by the former system, including explainability, contextual understanding, and granular comparison of mutual funds.
The process under the lens of the ontological framework\cite{Chen:2019:CGF} is outlined in \textit{Figure} \ref{fig:cs-1}. The baseline workflow that investors encounter is characterized by high interaction costs, as investors navigate a sizable candidate space to compare the intricacies of different mutual funds, where the trade-offs are complex. Both papers\cite{Yan2025VisualAnalysis} and\cite{Yan2025FundSelector:Selection} highlight the difficulty of mutual fund selection due to the large number of funds and the heterogeneity between investor preferences, while the latter research further shows that existing platforms provide many different attributes or basic return-based rankings without allowing investors to translate their investment strategy into practical selection criteria. This can be abstracted into the first \textbf{symptom} (Int-High-Ct), since the cost of investigating, comparing, recalling, and deciding is simply too high. A credible \textbf{cause} is insufficient statistical and algorithmic compression: the platforms offer too many alternatives or poor filtering and summarisation tools, reflected in the \textbf{causes} (Alg-Low-AC, Stat-Low-AC). A response to this is the first \textbf{remedy} highlighted in the earlier system: increasing algorithmic compression (Alg-High-AC) utilizing a mixed-initiative ranking model. By converting the selection issue into pairwise comparisons and injecting human input via learned preference weights, the classifier, enhanced by iterative investor feedback, narrows the initial candidate space to a more manageable ranked list of funds. Following multiple rounds of refinement, the quantitative MAP and NDCG assessments demonstrate that the learned model can be useful to support the preference selection process, thus substantially reducing cognitive costs.

However, the first \textbf{remedy} creates a \textbf{side effect} that receives limited attention in both papers but becomes apparent through abstract reasoning: once ranking is delegated to a machine learning model, investors are equipped with limited tools to examine the factors impacting the rankings, possibly resulting in blind trust or unverifiable suspicion. This results in a second \textbf{symptom} (Alg-High-PD), materializing as imperfect model performance and uncertainty about recommendation quality. The \textbf{cause} of this \textbf{symptom} may stem from insufficient training data and user feedback compared to the size of the total candidate space, as well as temporal changes in the information space (Stat-High-PD), which may be addressed by the second \textbf{remedy} retraining the ML model (Stat-Low-PD), which in turn still results over time in \textbf{side effects} (Alg-High-PD, Int-High-Ct). The performance of mutual funds is strongly dependent on changing market conditions, sector rotations, and changes in fund managers, so the information space itself changes significantly over an extended period. Although the weights of a trained model may be reasonable for one period, they may become less reliable over time. This explains why even a sensible machine learning algorithm \textbf{remedy} may still lead to significant residual distortion in deployment.

Subsequently, the 2025 FundSelector workflow can be viewed as a third major \textbf{remedy} for the residual distortion and explainability issues (Alg-High-PD, Int-High-Ct). The core contribution is not only additional visualization but also access to original, more contextual information via the coordinated views, thereby reducing the excessive compression of the algorithm (Alg-High-AC). The investment preference view presents and validates the current preference weights, the Market Overview provides temporal stock, bond, and sector market context, the Fund List provides context to ranking outcomes, the Fund Indicator view allows for multidimensional comparison, the Fund Manager view provides further investment style insights, and the Fund Comparison view provides temporal mutual fund view details via elastic trend charts. Furthermore, the rank-informed bipartite contribution bar chart provides insight into the positive and negative effects of the ranking process, thereby increasing the model's transparency. The latter 2025 work addresses Alg-High-AC with Vis-Low-AC, making reasoning more transparent to investors through FundSelector. This approach contrasts with the basic prototype of the earlier research and reduces the information hidden within the classifier. The new \textbf{remedies} may \textbf{cause} further \textbf{side effects}, including high interaction and learning costs, particularly for ordinary inexperienced investors (Int-High-Ct), in the form of interface complexity, navigation difficulty, and higher response times, which could be addressed in turn by progressive, personalized information disclosure (Vis-High-AC, Int-High-AC). Furthermore, this confirms the central claim of the framework that workflow optimization is iterative.

However, compared to previous iterations, the improved system is more efficient, specifically because it does not replace complex human judgment with machine learning alone, but instead supports decision-making through the strengths of algorithmic preference learning and human-driven analysis utilizing visualization.  At a higher level of abstraction, this case study illustrates an issue between machine learning development and deployment workflows. During the development process, the accuracy of the model is improved through relabeling, retraining, validation, and testing, while in deployment the model has to operate on evolving data, relying on the soft knowledge of the investor to detect outdated weights and discrepancies, i.e., to monitor the changing information space.

%% file: Appendix/2.CaseStudy-TextSimilarityDetection.tex
This case study is based on the Text Similarity Detection work of Abdul-Rahman et al.~\cite{abdulrahman:2017:CGF}. The work was part of the JISC/NEH (III) Digging into Data Challenge program. The team consists of visualization scientists and domain experts working in literary studies, intellectual history, and digital humanities, distributed across three continents: the UK, the US, and Australia. In this work, we use a novel visual analytics approach to identify commonplaces in 18th-century literary and print culture.
 
Commonplace is a thematic collection of quotations and similar passages intended for future recall and reuse. In other words, two similar sequences in different settings can be considered commonplaces. The 18th century can be viewed as one of the last periods in a long tradition of `commonplace cultures' stretching from Antiquity through the Renaissance and Early Modern eras, where commonplaces often mirror the `social network' of that time. Essentially, commonplaces likely reveal who liked whom in the 18th century; if two people liked each other, they tended to copy more from one another.
 
Detecting similarity between texts is a common task in text mining that domain experts have studied extensively, often using automated methods such as machine learning models. Because similarity measurement typically relies on multiple metrics, some of which are sensitive to subjective interpretation, a generic machine-learning-based detector often struggles to balance the roles of different metrics across the semantic context of a specific text collection. Although these methods have seen considerable success, they also produce many false positives that can be time-consuming to eliminate.

\begin{figure*}[ht]
    \centering
    \includegraphics[width=\linewidth]{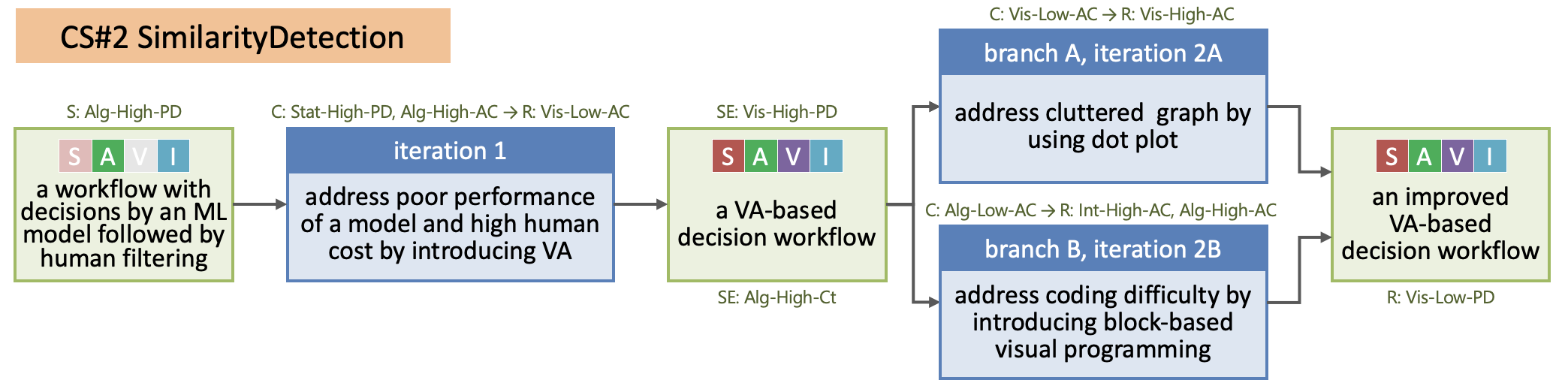}
    \caption{\textbf{CS\#2 SimilarityDetection}: An updated workflow for optimizing workflows (WF4OWF) for the SimilarityDetection case.}
    \label{fig:cs-text-smilarity}
\end{figure*}

\textit{Figure}~\ref{fig:cs-text-smilarity} shows the process under the lens of the ontological framework [1]. Here we identify the first \textbf{symptom}: domain experts have experience with machine learning models, but the current workflow with these models produces too many false positives. This \textbf{cause}s a time-consuming process to eliminate false positives, making it harder for domain experts to focus on the actual cases. The information is being compressed too quickly and excessively, leading to high entropy and high Alphabet Compression (AC). A possible \textbf{remedy} is to use visualizations to represent the results as alternatives to machine learning models, especially since we have previously noted that these models compress information too much and too quickly. A \textbf{side effect} of this approach is to determine which visualization method – a bipartite graph or a 2D pixel map – best represents commonplaces.
 
The next \textbf{symptom} is that when we visualize texts as a bipartite graph, the distortion becomes too high, making it difficult to see patterns where the information cost is low but the display cost is high because the AC is insufficient. This \textbf{cause}s a cluttered visualization with too few pixels, and the bandwidth cost also becomes too high. A possible \textbf{remedy} is to use a 2D pixel map or matrix to display text or paragraphs. However, a \textbf{side effect} of this approach is increased cognitive load due to the added complexity of the 2D pixel map or matrix visualization.
 
Our third \textbf{symptom} involves choosing which algorithms domain experts should use to identify text similarities in paragraphs. This \textbf{cause}s the problem that, when many algorithms are available, both humans and machines struggle to select and combine them, increasing complexity. A possible \textbf{remedy} is to add interactions to our visual analytics pipeline to reduce this complexity. A \textbf{side effect} is that building the algorithms and pipeline requires a learning process, but the learning costs are justified.

%% file: Appendix/3.CaseStudy3BookTrade.tex
\subsection{Initial Case Study}
This case study is based on the development of \emph{BookTracker}~\cite{xing:2024:TVCG}, a visual analytics system designed to support historians in studying the provenance and circulation of early printed books. The work was conducted through a close collaboration between visualization researchers and historians specializing in the history of the book trade. The data used in this work is derived from the Material Evidence in Incunabula (MEI) database~\cite{database_mei_2015}, which records the provenance evidence and institutional holdings of books printed in Europe during the early age of print.

\begin{figure}[ht]
    \centering
    \includegraphics[width=\linewidth]{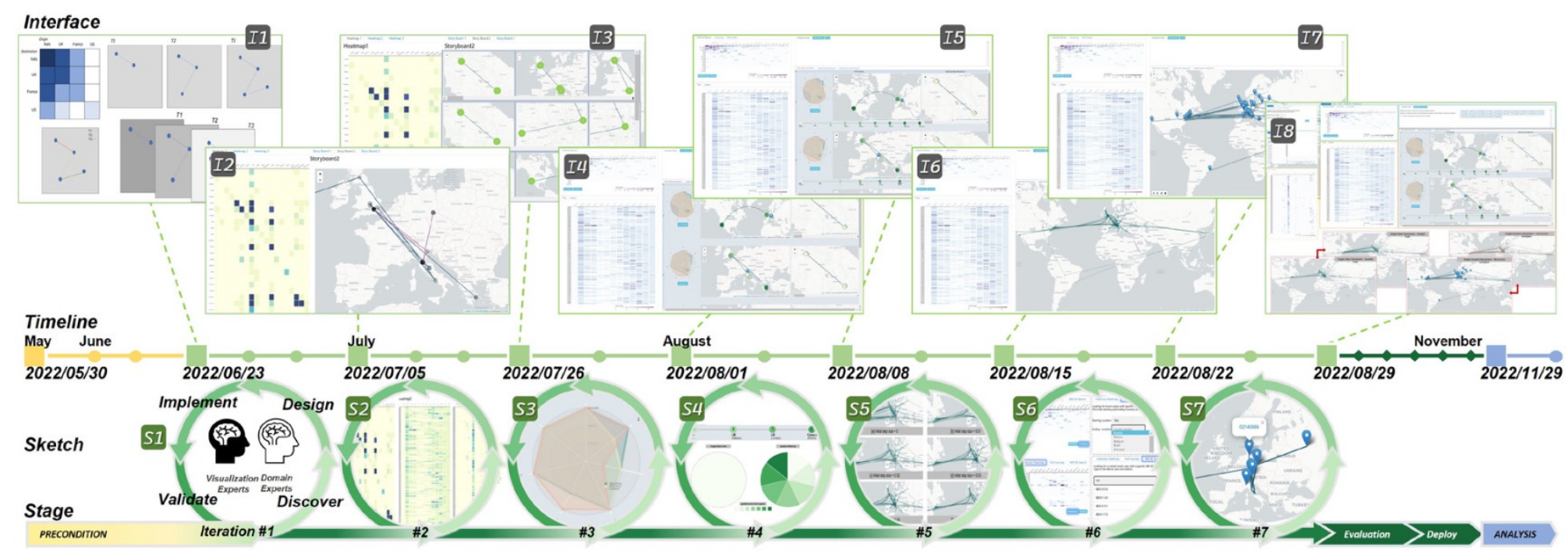}
    \caption{\textbf{CS\#3 BookTrade}: The design iterations of \emph{BookTracker}.}
    \label{fig:cs3-book-trade-design}
\end{figure}

The entire design process can be summarized as seven design iterations following repeated \textit{design–implement–feedback} loops, as illustrated in \textit{Figure}~\ref{fig:cs3-book-trade-design}. From the overall design and collaboration process, we extracted key domain requirements as follows:
\begin{itemize}
    \item [R1] Search and visualize the provenance history of one or multiple books.
    \item [R2] Visualize the provenance paths on the geo-map.
    \item [R3] Present the provenance history through animation.
    \item [R4] Visualize the detailed statistics of the provenance records.
    \item [R5] Support the detection of similar patterns among book records. 
\end{itemize}

These requirements were not identified in one go. Instead, throughout the collaboration, we observed that domain experts do not always have clearly defined analytical problems in mind, nor are they always familiar with articulating their needs in computational terms~\cite{xing:2024:TVCG}. Consequently, the design process evolved largely through discussions during feedback meetings. Importantly, the visualization tasks addressed in each iteration were not always refinements of previous designs; rather, new analytical directions often emerged, leading to branching problem explorations.

\begin{figure*}[ht]
    \centering
    \includegraphics[width=\textwidth]{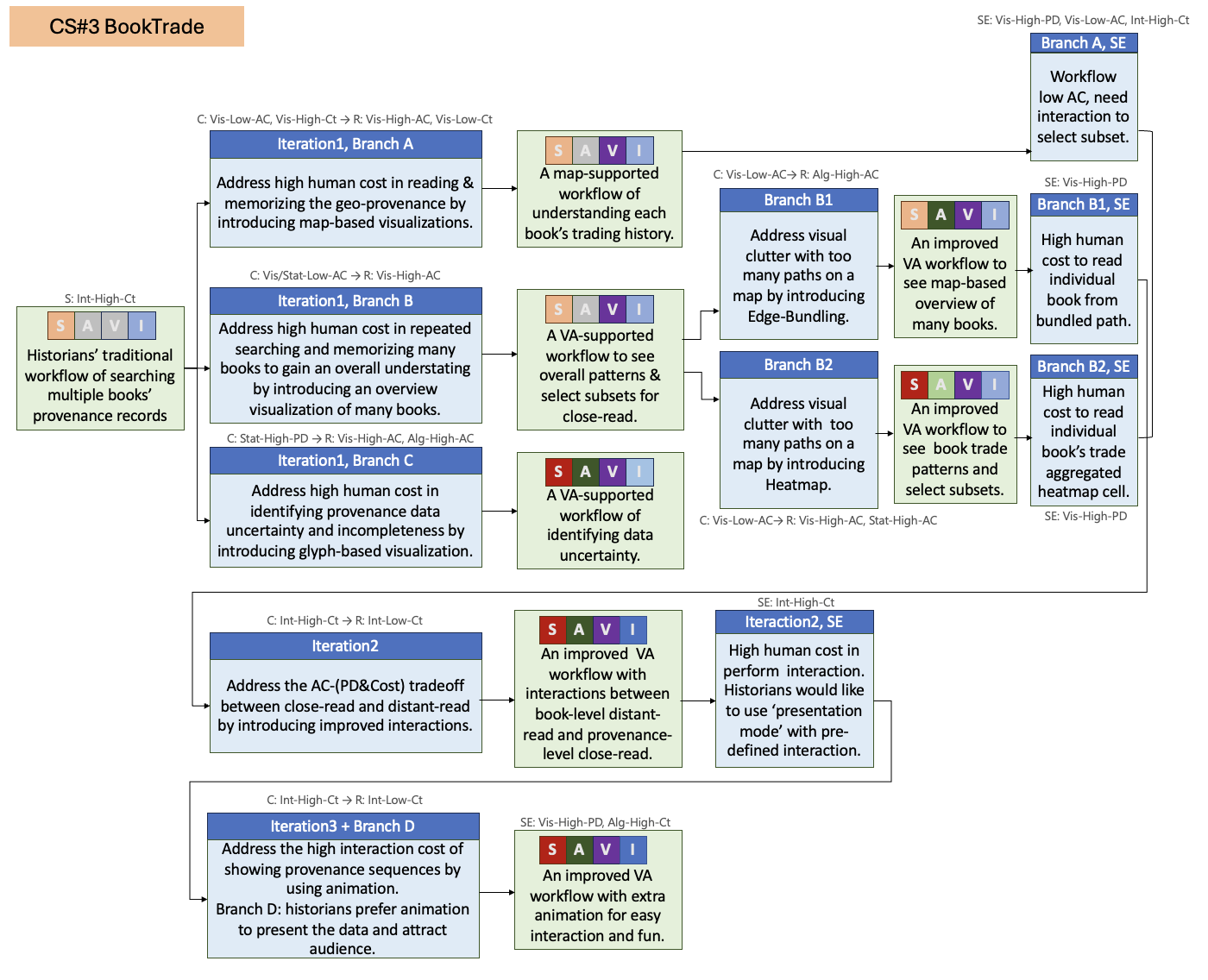}
    \caption{\textbf{CS\#3 BookTrade}: A workflow for optimizing workflows (WF4OWF) for the BookTrade case. Note that in the main text, several branches of this workflow are combined into a single iteration in the interest of brevity. The decomposed version is depicted here.}
    \label{fig:cs3-book-trade-WF4OWF}
\end{figure*}

Here, we revisit our iterative design process retrospectively using the \textbf{SCORE} framework. The Workflow for Optimizing Workflows (WF4OWF) is illustrated in \textit{Figure}~\ref{fig:cs3-book-trade-WF4OWF}. Taking into account the baseline approach used by historians to search provenance records, the interface is heavily text-based, with no visualization, no algorithmic assistance, limited interaction, and minimal statistical support. Users must select each book entry from the catalog individually and open a new tab to view its provenance records. The key \textbf{symptom} and \textbf{cause} here is that it is cognitively costly for users to memorize and compare the provenance histories of multiple books simultaneously, while the search interaction itself is tedious and inefficient. Information compression in the current interface is low.

To optimize this workflow of searching and memorizing provenance information, the first \textbf{remedy} is to introduce geo-based visualizations to represent provenance data as an alternative to textual records. In this representation, the provenance locations of a book are compressed into points (cities) on a map, while the movement sequence is represented as line segments connecting these points. This allows users to view the provenance histories of multiple books simultaneously within a single visual space. However, a \textbf{side effect} of this approach is that when the number of books becomes large, the visualization may become cluttered. 

A second \textbf{remedy} is to introduce an overview visualization that supports pattern identification throughout the entire dataset and enables users to select subsets of books more efficiently. Nevertheless, this approach may again result in visual clutter when too many provenance paths are displayed simultaneously (\textbf{side effect}). To address this, two alternative \textbf{remedies} can be considered. 1) The Edge-Bundling algorithm can be introduced to reduce visual clutter by grouping similar paths together. While this reduces display complexity, a potential side effect is that the resulting curved lines may introduce information distortion, making it harder for users to determine which paths correspond to individual books. 2) A heatmap representation can replace the paths on the geographical map by displaying only the frequency of transfers between locations. This approach mitigates map clutter and compresses the provenance information, but introduces a different \textbf{side effect}: the detailed provenance information of individual books becomes distorted.

We observe that the improved visual analytics workflow introduces a fundamental trade-off between close- and distant-reading of provenance information (\textbf{symptom}). In other words, this reflects a trade-off between higher Alphabet Compression (AC) and the resulting increase in Potential Distortion (PD) and cognitive cost (cause). To address this issue, additional interactions can be introduced to bridge summary visualizations and detailed views (\textbf{remedy}), allowing users to move seamlessly between overview and detail. However, the \textbf{side effect} is that users may need to perform multiple interaction operations, increasing human labor cost.

A further \textbf{remedy} is to introduce animation, which can automate sequences of interactions and allow users to observe the system revealing the provenance dynamically. Interestingly, these analyzes echo the requirements from our real-world collaboration: the domain experts explicitly requested animation as an engaging way to present their data. But here it can also serve as a means of reducing the human effort required to manually perform multiple interaction steps.

\subsection{Update}
\begin{figure*}[ht]
    \centering
    \includegraphics[width=\textwidth]{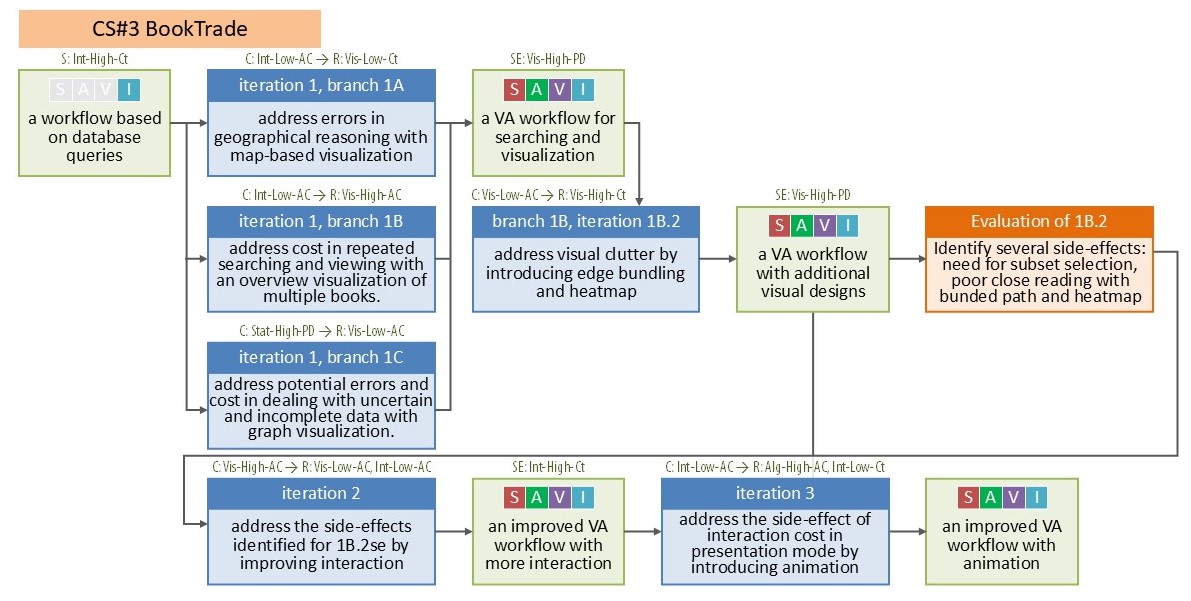}
    \caption{\textbf{CS\#3 BookTrade}: An updated workflow for optimizing workflows (WF4OWF) for the BookTrade case.}
    \label{fig:cs3-book-trade-WF4OWFb}
\end{figure*}

\textit{Figure} \ref{fig:cs3-book-trade-WF4OWFb} was updated following group discussions. 

%% file: Appendix/4.CaseStudy4-DataVirtualization.tex
\subsection{Workflow Overview}

\begin{table*}[tbh]
\centering
\caption{\textbf{CS\#4 DataVirualization}: Abstract reasoning for symptoms, causes, remedies, and side-effects associated with a data virtualization workflow applied using SCORE ontological framework. Rows 2 and 3 trace the side-effects of the remedy introduced in Row 1. 
}
\label{tab:data-virtualzation}
\renewcommand{\arraystretch}{1.35}
\setlength{\tabcolsep}{3pt}
\scalebox{0.78}{%
\begin{tabular}{|p{3.65cm}|p{3.65cm}|p{3.65cm}|p{3.65cm}|p{3.65cm}|p{3.65cm}|}
    \hline
    \textbf{Symptoms} &
    \textbf{Abstract Reasoning (of symptoms)} &
    \textbf{Possible Causes} &
    \textbf{Abstract Reasoning (cause $\rightarrow$ prescription)} &
    \textbf{Remedies} &
    \textbf{Abstract Reasoning (side-effects)} \\
    \hline
    ML developers repeatedly perform identical data wrangling tasks (e.g., selection and processing) across experiments and across team members. &
    (I, A, S) Low-AC &
    No shared, compressed representation of prior transformation efforts exists; every developer independently processes raw data from scratch. &
    (I, A, S) Low-AC $\rightarrow$ (I, A, S) High-AC &
    Centralize and store all previously computed transformation outputs so that subsequent developers can reuse them without re-implementation. &
    A High-Ct \\
    \hline
    Storing all preprocessed data permutations as explicit physical copies consumes excessive storage space (side effect of row 1). &
    A High-Ct &
    Explicit duplication: every unique transformation output is materialized as a separate physical dataset. &
    A Low-AC $\rightarrow$ A High-AC &
    Data virtualization: replace explicit copies with virtual datasets. &
    A Low-Ct \newline I High-Ct (at query time) \\
    \hline
    Virtualized data are not transparent to developers who did not create them: retrieving the processing history / auditing is not easy (side effect of row 2). &
    I High-AC &
    Lost provenance &
    I High-AC $\rightarrow$ V Low-AC &
    Provenance visualization: expose the complete data lineage (e.g., source datasets, transformation functions, parameters, etc.) through a visualization interface. &
    I Low-Ct \\
    \hline
\end{tabular}%
}
\end{table*}

In traditional machine learning (ML) workflows, developers perform numerous experiments that require extensive data wrangling, e.g., feature extraction, normalization, data merging, etc., and partitioning datasets into training, validation, and testing sets. Each developer independently writes programs that produce explicit physical copies of the transformed datasets at every stage of the pipeline, as illustrated in Figure 2 (a) in~\cite{Khan:2025:SCC}. This practice results in massive storage redundancy, inconsistent and duplicated transformation efforts across team members, and a loss of data provenance whenever intermediate datasets are overwritten or deleted during iterative experimentation.

Khan et al. proposed Data Virtualization~\cite{Khan:2025:SCC} to address these inefficiencies by replacing explicit data copies with virtual datasets: light-weight specifications that record the path links to the source data and the transformation functions required to reproduce any derived dataset on demand. As shown in Figure 2 (c) in~\cite{Khan:2025:SCC}, a centralized data virtualization service processes data queries, traces the virtualization graph back to the original explicit datasets, executes the required transformations, and delivers the materialized data to ML workflows without creating intermediate files. This solution eliminates storage redundancy, creates consistent transformation pipelines, and preserves the complete data lineage as a virtual dataset.

\subsection{Ontological Cost-Benefit Analysis}
\begin{figure*}[ht]
    \centering
    \includegraphics[width=\linewidth]{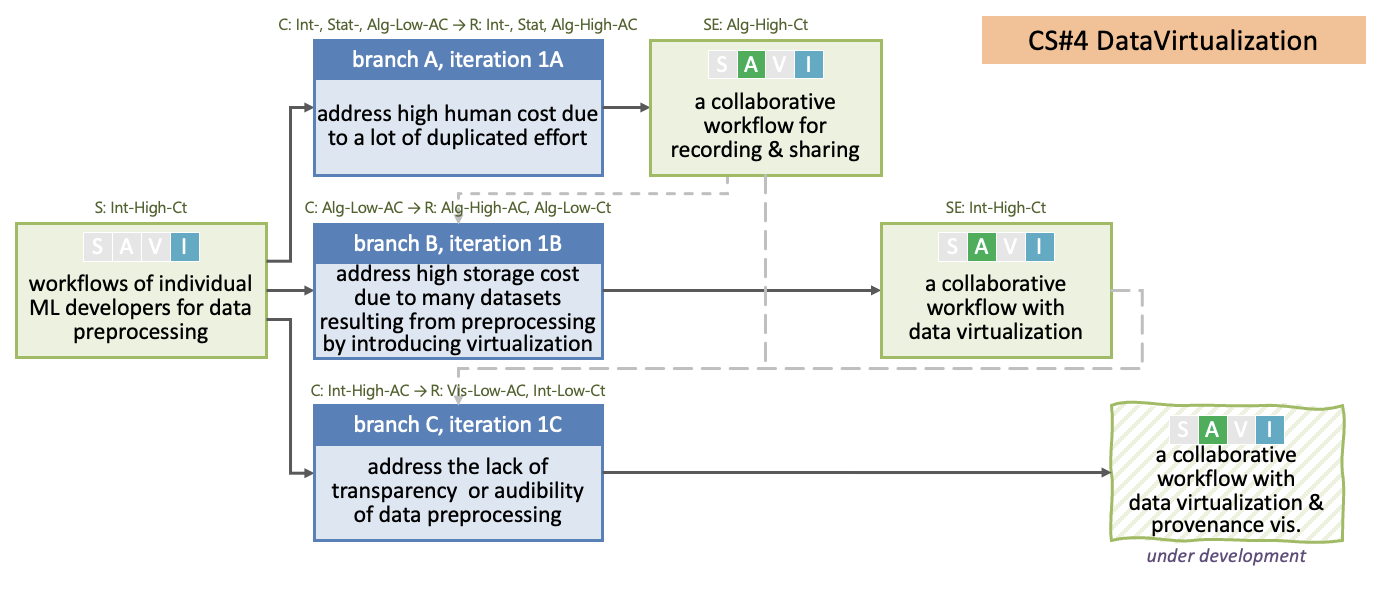}
    \caption{\textbf{CS\#4 DataVirualization}: An updated workflow for optimizing workflows (WF4OWF) for the DataVirtualization case.}
    \label{fig:datavirtualization}
\end{figure*}

Table~\ref{tab:data-virtualzation} discusses how the SCORE framework supports structured and iterative design optimization of the data virtualization workflow. The Workflow for Optimizing Workflows (WF4OWF) for the DataVirtualization case is depicted in \textit{Figure}~\ref{fig:datavirtualization}.

\vspace{2mm} \noindent 
\textbf{Primary symptom and diagnosis.}  
The principal \textbf{symptom} in conventional ML workflows is that developers continually repeat the same data-wrangling tasks from scratch. From an information-theoretic perspective, this represents excessively low alphabet compression (AC) across all three abstract components: interaction (I), algorithm (A), and statistics (S), because no shared, compressed representation of prior transformation effort is available. Every developer must independently sift through raw data and process derivations that may already have been performed by colleagues. 
The root \textbf{cause} is therefore diagnosed as (I, A, S) Low-AC. The \textbf{remedy} is to store transformation outputs centrally so that downstream developers can reuse them. Storing pre-processed datasets increases AC across I, A, and S (denoted as (I, A, S) High-AC) by creating a shared dataset that can be queried on demand. This \textbf{remedy} reduces the interaction costs borne by individual developers; however, materializing all possible transformation permutations as explicit physical copies imposes an algorithmic storage cost (A High-Ct). 

\vspace{2mm} \noindent 
\textbf{Storage overhead.} 
This cost is a direct consequence of explicit duplication of the derived dataset (A Low-AC). Each derived dataset is stored in a separate physical file, increasing storage requirements with each new transformation variant explored during experimentation.
In some cases, for example, when a long time series is segmented using a sliding window, the size of the derived dataset can approach a multiple of the original source. Data virtualization resolves this storage \textbf{side-effect} by storing only lightweight specifications, such as paths to source data and identifiers for transformation functions, rather than the materialized outputs. This shift from physical to virtual representation increases algorithmic AC (A High-AC) by collapsing multiple redundant files into a single specification entry. 
The cost introduced is a marginal computational overhead (Low-Ct) incurred when a virtual dataset is converted to a materialized dataset by executing the required transformations at query time (High-Ct).
However, this overhead is insignificant in practice, especially considering the affordability of memory and processing power.

\vspace{2mm} \noindent 
\textbf{Loss of provenance and transparency.} 
The introduction of virtual datasets creates a new issue: developers who use these datasets produced by others cannot see the history of transformations applied.
The graph representing data path links, transformation functions, and parameter choices in the virtualization pipeline is difficult to interpret.
This situation highlights a high level of interaction complexity (I High-AC), as developers have lost access to essential provenance information needed to understand, verify, or reproduce data derivation.
However, it is important to note that the virtualization infrastructure does not discard provenance information. 
In contrast, the complete data lineage is structurally preserved within the specifications of each virtual dataset, detailing all path links and transformation functions. 
The appropriate \textbf{remedy} is therefore to lower the AC through provenance visualization (V Low-AC). By visualizing the transformation graph as an interactive, navigable structure, developers can audit, verify, and reproduce data transformations, resulting in low interaction cost (I Low-Ct) throughout the ML development lifecycle. 

%% file: Appendix/5.CaseStudy.tex
\begin{table*}[!tb]
\centering
\caption{\textbf{CS\#5 SubspaceAnalysis}: Summary of the ontological framework applied to the design process of the multi-dimensional pattern exploration technique and application. 
}
\label{tab:mdspe-table}
\renewcommand{\arraystretch}{1.35}
\setlength{\tabcolsep}{3pt}
\scalebox{0.78}{%
\begin{tabular}{|p{5.6cm}|p{5.6cm}|p{5.6cm}|p{5.6cm}|}
    \hline
    \textbf{Symptoms} &
    \textbf{Possible Causes} &
    \textbf{Remedies} &
    \textbf{Side-Effects} \\
    \hline
    Dashboards require a-priori knowledge and interactive search to find subspaces & Exponential combination of possible subspaces, too much to visualize at once & Exploiting the curse of dimensionality + default parameters to show all subspaces simultaneously & All relevant subspaces are presented in a single graphic – may overwhelm the user \\
    \hline
    Fully automated approaches do not consider a-priori knowledge of the user and can find many invaluable, uninteresting correlations &
    Curse of dimensionality dramatically decreases the size of subspaces and makes deviations more easily statistically significant -> spurious correlations &
    Allowing to use of statistics and providing it as an overlay, allowing the user to remove unwanted correlations & Not filtering algorithmically produces visual noise \\
    \hline
    Cross-filter dashboards do not easily allow comparison of subspaces & Too many possible subspaces and application-dependent metrics necessary for comparison & Focus on correlation measures (\& deviation) to simplify visualization & Visual clutter \\
    \hline
    User overwhelmed by visual complexity & Highly dense pixel visualization with visual noise & Automatically sorting rows to highlight redundant patterns, allow the user to order columns, and filter by columns, guidance features showing the user possibly interesting subspaces & Increasing the complexity of the user interface \\
    \hline
\end{tabular}%
}
\end{table*}

\begin{figure*}[ht]
    \centering
    \includegraphics[width=\linewidth]{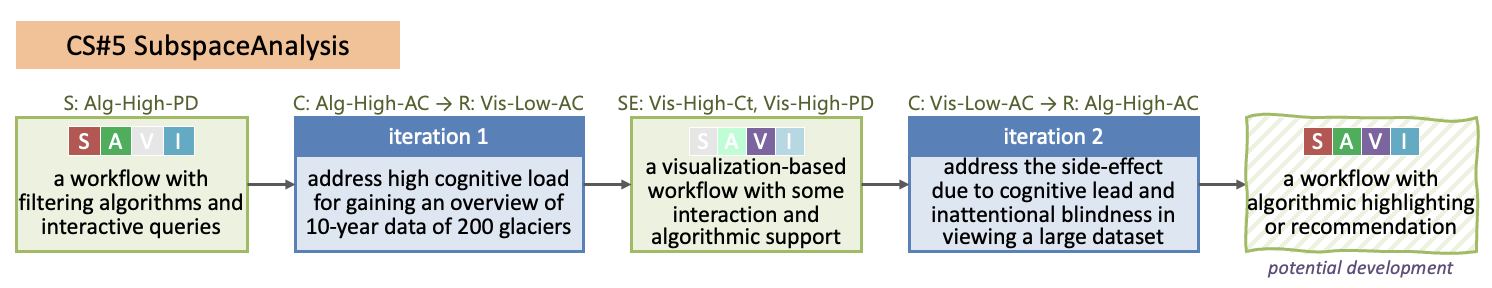}
    \caption{\textbf{CS\#5 SubspaceAnalysis}: An updated workflow for optimizing workflows (WF4OWF) for the SubspaceAnalysis case.}
    \label{fig:subspaceanalysis}
\end{figure*}

This case study provides a post-hoc analysis of Chen's and Ebert's methodology\cite{Chen:2019:CGF} to the multi-dimensional pattern exploration technique and the prototype developed by Jentner et al.\cite{jentner2023}. The Workflow for Optimizing Workflows (WF4OWF) for the SubspaceAnalysis case is illustrated in \textit{Figure}~\ref{fig:subspaceanalysis}. The subspace analysis of structured data is relevant for tasks such as cohort analysis of patient histories and pharmaceutical research. The data in question are structured data in the form of sets, sequences, trees, or graphs, and are paired with discrete attributes such as gender, age, and patient medical conditions. The goal is to find correlations between sub-structures of the structured data and the attributes based on co-occurrence distributions. An example of a cohort analysis would be that a certain subset of patients with similar medical histories correlates with attributes of men over the age of 60 years, having type 2 diabetes and hypertension. This is challenging as both the sub-structures as well as combination of attributes grow exponentially. Moreover, the interestingness of correlations based on co-occurrence distributions is application-, user-, and task-dependent and cannot be easily automated to filter the computational space.

A dashboard with cross-filtering features allows the user to filter the data by multiple attributes, running algorithms, and statistics only on the user-filtered subset (interaction first approaches). Therefore, the \textbf{symptoms} of such approaches are high interaction cost and high alphabet compression, but they simultaneously introduce high potential distortion, as users cannot execute the algorithm for every possible filter setting.
Algorithm-first approaches that execute over the entire dataset require a-priori knowledge of the user to identify potential correlations, thus increasing their alphabet compression. Inappropriate parameter settings for algorithms \textbf{cause} an explosion in visualization costs, as an exponential number of results is generated and must be assessed by the user.

Jentner et al.'s approach is algorithm-first, but it introduces an algorithm that eliminates redundant data by linearizing one of the exponential dimensions and significantly reducing the second dimension. Although this introduces some algorithmic cost, it eliminates interaction cost, as a single-pixel-based visualization can be rendered. This visualization is difficult to interpret and navigate, as the information density is high (high visualization cost).

As a \textbf{remedy} to the high visualization cost, interactive selection was introduced, allowing users to select specific rows in either table and have the visual links highlighted, rather than displaying all links simultaneously. The application introduced zooming, panning, and tooltips to support visual navigation. These features slightly increase interaction cost but significantly reduce the risk of inattentional blindness by allowing the user to focus on manageable subsets of the pixel space. Furthermore, the application allows one to switch normalizations and adjust the color of the pixels, enabling users to reveal different visual patterns from the same underlying data. This corresponds to reducing the potential distortion of the visualization to match the user's current analytical question. To further improve this, tables can be reordered based on user-defined criteria (e.g., interestingness measures such as support, LIFT, or confidence), placing regions of potential interest at the top or bottom and supporting navigation. Rows outside user-defined boundaries can be removed, reducing visual complexity. This complements sorting by narrowing the displayed data to regions of interest.

The approach inverts the conventional workflow of starting with attributes to filter data. Instead, it presents all sub-structures and their co-occurrences simultaneously, requiring users to adopt a new mental model. Combined with the accumulated interactive features, this steepened the learning curve for novice users, who were unsure how correlation patterns would manifest in the visualization.
As a \textbf{remedy}, features were introduced that allow users to search for specific structured data patterns (e.g., "the car drives this road and then that one"), filter the table directly, and support hypothesis testing. This reduces the interaction cost by providing a familiar entry point into the otherwise unfamiliar visual representation. In addition, the system lists potential correlation patterns of interest. Clicking on a suggestion highlights the relevant rows and automatically pans and zooms the canvas to the correct location. This serves a dual purpose: to reduce the cost of initial exploration and to help users learn how various correlation patterns are visually represented in the pixel space, thereby reducing the long-term cognitive cost.

%% file: Appendix/6.CaseStudyPromptsforLLMs.tex
\subsection{Initial Case Study}
This case study is a post-hoc analysis of Hao et al.’s FinFlier\cite{Hao2025}, a visual analytics system that produces graphical overlays to visualize financial narratives. It addresses a workflow issue in financial analysis: textual data and charts are commonly examined side by side, increasing the cognitive burden for readers to compare values, trends, turning points, or other patterns related to textual and graphical data. FinFlier tackles this issue by combining a knowledge-grounded LLM for text-data binding with a graphical overlay module that maps the retrieved narrative to appropriate visual annotations. The method is based on a survey of 1,752 layered financial and academic charts, identifying common narrative structures, overlay methods, and correspondence patterns\cite{Hao2025}. Hao et al.\cite{Hao2025} utilize output constraints, chain-of-thought reasoning, and dynamic few-shot prompting to detect subjects, trend patterns, and numerical values and connect them to tabular data\cite{Hao2025}, resulting in automated text-data binding and graphical overlays.

\begin{figure*}[ht]
    \centering
    \includegraphics[width=\linewidth]{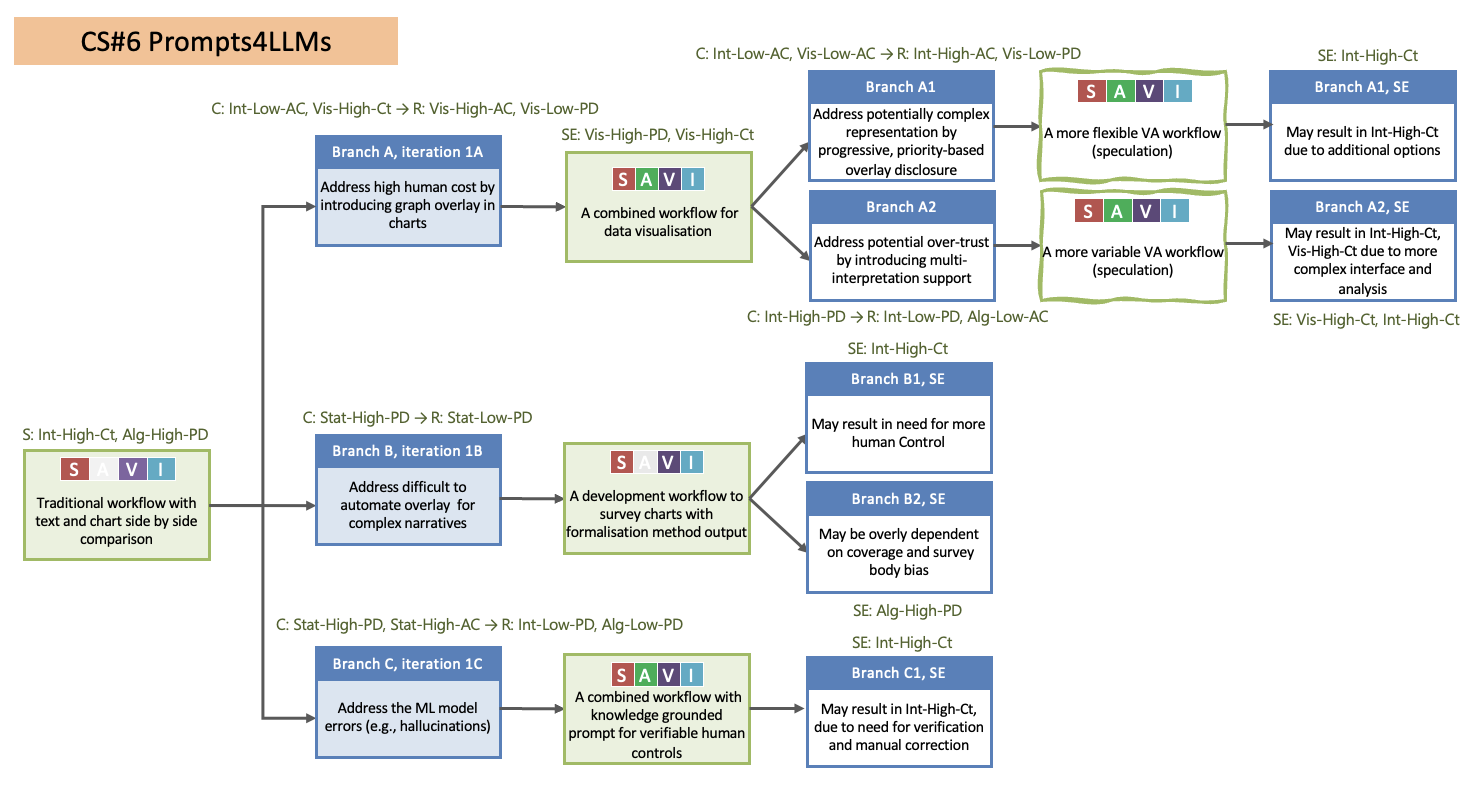}
    \caption{\textbf{CS\#6 Prompts4LLMs}: A workflow for optimizing workflows (WF4OWF) for the Prompts4LLMs case. Note that this version provides an alternative interpretation compared to the concise version in the main text in the \textit{Figure} \ref{fig:CaseStudyWFs}, splitting symptoms and following reasoning into distinct branches.}
    \label{fig:Prompts4LLMs}
\end{figure*}

Applying the Chen and Ebert’s method\cite{Chen:2019:CGF}, \textbf{symptoms}, \textbf{causes}, \textbf{remedies} and analysis of \textbf{multi-order effects} can be subdivided into branches, as outlined in \textit{Figure} \ref{fig:Prompts4LLMs}. In the baseline workflow, users need to inspect financial narrative statements in one view and analyze a chart in the other, while mentally establishing a connection between them, which results in high interaction cost (Int-High-Ct) due to split attention and slow comprehension. The most likely \textbf{causes} are the disconnection between the text and the visualization, as well as the insufficient structure for encoding key narrative-data patterns (Int-Low-AC, Vis-High-Ct). As a first \textbf{remedy}, graphical overlays are directly embedded in the chart, including highlights, labels, descriptions, and trend lines, thus externalizing reasoning and encoding narrative meaning as part of the visualization, which naturally maps to Vis-High-AC and Vis-Low-PD. However, this leads to \textbf{side effects} that the original paper only partially addresses, such as a denser, more complex representation. Although overlays substantially decrease the cognitive cost of identifying key patterns, they risk clutter, occlusion, and competition among elements within the visualization (Vis-High-PD, Vis-High-Ct). These could be addressed by introducing progressive, priority-based overlay disclosure (Int-High-AC, Vis-Low-PD), but this may result in Int-High-Ct due to additional actions required to access optional annotations. Another \textbf{multi-order effect} is that users over-trust the overlay narrative as an authoritative interpretation of the data (Int-High-PD), even though it is only one among multiple plausible alternatives. A possible \textbf{remedy} could be multi-interpretation support with uncertainty and alternatives in the form of a toggle between possible annotations (Int-Low-PD, Alg-Low-AC), but this may result in Int-High-Ct or Vis-High-Ct, as multiple alternatives make the interface and analysis more complex.
The second \textbf{symptom} concerns the difficulty of producing useful graphical overlays for complex financial narratives (Alg-High-PD). Initially, there is no correspondence between financial narrative types and graphical overlay patterns, so an automated system would have to address the mapping issue arising from underspecified inputs (Stat-High-PD). The \textbf{remedy} is a development workflow based on a survey of 1,752 real-world charts, resulting in a formalization of common narrative structures, overlay types, and correspondence patterns (Stat-Low-PD). This \textbf{remedy} leads to dependence on coverage and the survey body's bias, as well as the need for rigorous human controls, since underrepresented financial narratives might be misrepresented (Alg-High-PD, Int-High-Ct). However, it represents a clear improvement to the baseline workflow.
The third \textbf{symptom} is the machine learning bottleneck: Alg-High-PD due to errors in identifying relational patterns between text and data, including hallucination, unit mismatches, or incorrect subject detection. Likely \textbf{causes} include Stat-High-PD and Stat-High-AC, as financial language is specialized, context-dependent, and can exhibit sparse labeled training data. FinFlier addresses this issue by using knowledge-grounded prompting that constrains LLM's output to verifiable structures and human controls (Int-Low-PD, Alg-Low-PD), demonstrating improved text-data binding accuracy compared to baseline LLMs\cite{Hao2025}. However, the adverse \textbf{side-effect} is Int-High-Ct in the later stages of the process during required verification and manual correction. FinFlier does not simply represent automation, but a reconfiguration of earlier workflows, thereby reallocating cognitive costs from manual authoring to control and correction functions. 

\subsection{Update}
\begin{figure*}[ht]
    \centering
    \includegraphics[width=0.9\linewidth]{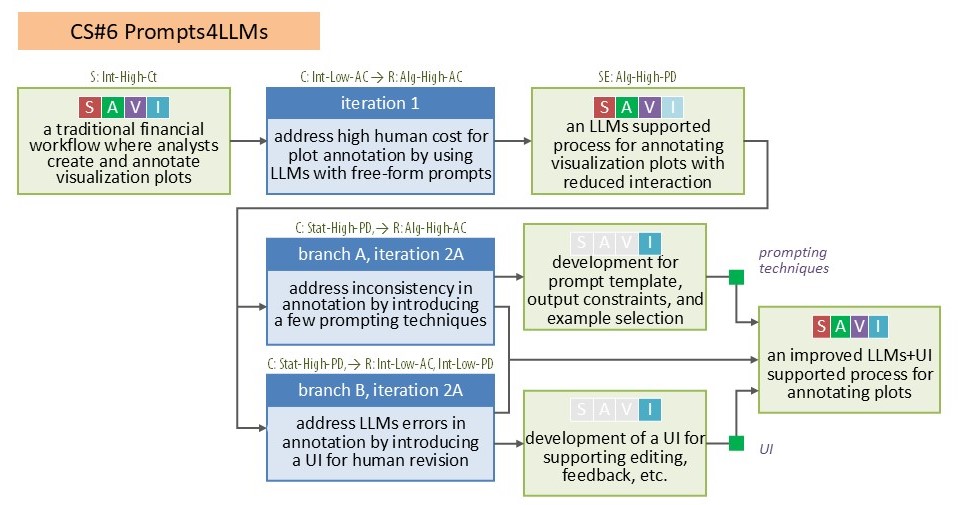}
    \caption{\textbf{CS\#6 Prompts4LLMs}: An updated workflow for optimizing workflows (WF4OWF) for the Prompts4LLMs case.}
    \label{fig:Prompts4LLMs2}
\end{figure*}

The following \textit{Figure} \ref{fig:Prompts4LLMs2} was updated on the basis of group discussions. 

%% file: Appendix/7.CaseStudy.tex
\begin{figure}[ht]
\centering
\begin{subfigure}{0.41\linewidth}
\includegraphics[width=\linewidth]{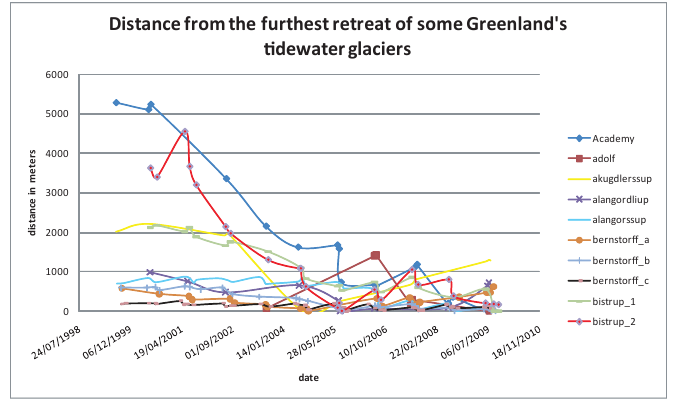} 
\caption{Time-series plot}
\label{fig:Timeseries}
\end{subfigure}
\begin{subfigure}{0.56\linewidth}
\includegraphics[width=\linewidth]{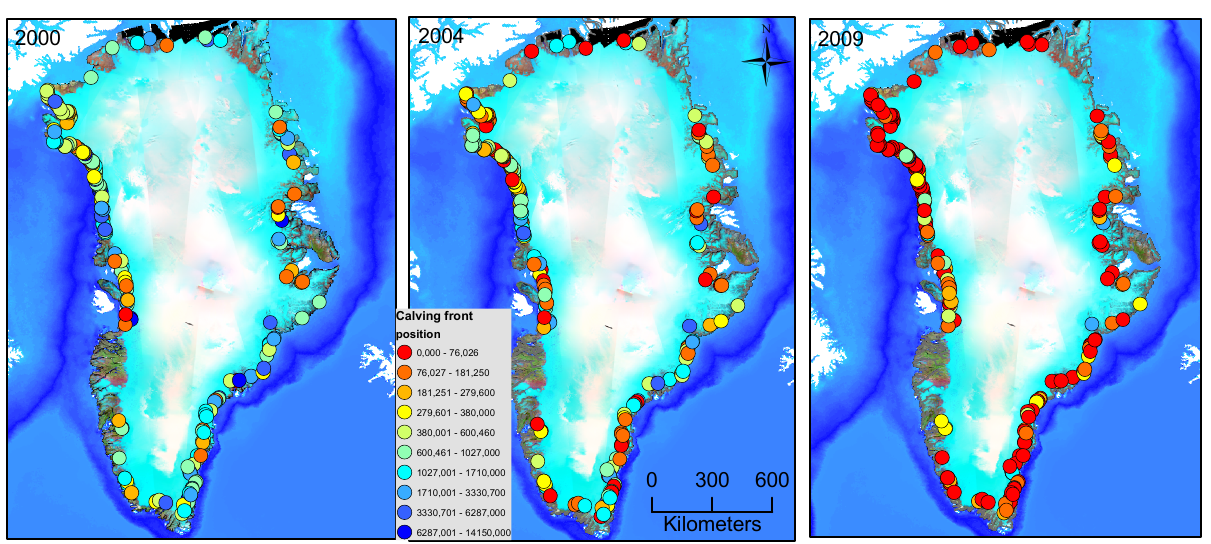}
\caption{Color coded evolution of the calving front positions}
\label{fig:ClavingFront}
\end{subfigure} \
\begin{subfigure}{0.8\linewidth}
\includegraphics[width=\linewidth]{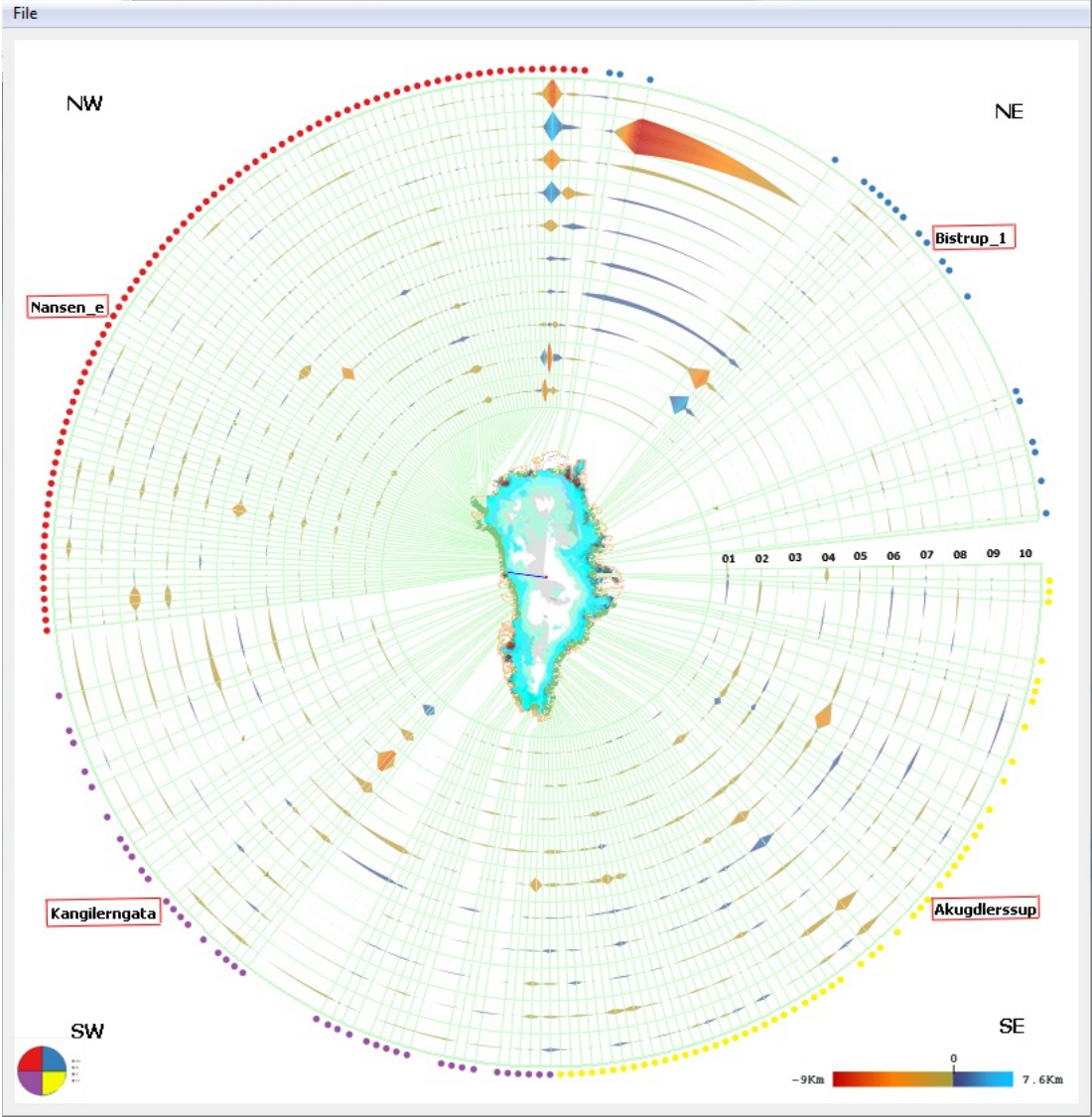}
\caption{Novel radial view design}
\label{RadialPlot}
\end{subfigure}
\caption{\textbf{CS\#7 GlacierMovement}: Traditional visualisations of glacier-front change: (a) time series of 10 calving glaciers; (b) color-coded glyphs showing relative frontal position over time, and sea surface temperature; (c) novel design: radial view of glacier-front change: area plot of 199 calving glaciers over 10 years, on a false-color Landsat image of Greenland..}
\label{fig:cs-7}
\end{figure}

This case study analyzes the visual analytics approach presented in Drocourt et al. \cite{drocourt-2011}. The work addresses the challenge of visualizing a ten-year record of seasonal and inter-annual changes in the frontal position of nearly 200 marine-terminating glaciers distributed along the Greenland coastline. The dataset was derived from Landsat satellite imagery and contains irregular time series describing glacier advance and retreat relative to a reference terminus position. The goal of visualization is to allow glaciologists to analyze spatial patterns and temporal changes in glacier behavior simultaneously, facilitating scientific understanding of glacier dynamics and their contribution to sea-level rise.
 Before the development of the proposed visualization, glaciologists relied primarily on two forms of visual analysis. The first approach consisted of time-series plots representing the relative frontal positions of individual glaciers over time (see \textit{Figure} \ref{fig:cs-7}-(a)). Although effective for examining temporal change of a single glacier, the plots did not support the preservation of spatial relationships (geographical position, proximity, etc.), making spatial comparisons difficult. The second approach used color coding in the form of map-based glyph visualizations where glacier changes were represented using color-coded symbols on geographic maps (see \textit{Figure} \ref{fig:cs-7}-(b)). Although these provide spatial context, it requires multiple juxtaposed views to analyze temporal variation, moreover it is prone to visual clutter when several glaciers are displayed. Furthermore, empirical studies have shown how the evaluation of temporal changes in color-coded pixel-based maps imposes a high cognitive load on human observers, leading to reduced accuracy and slower response times \cite{borgo-2010}.

%\begin{figure}[ht]
%    \centering
%    \includegraphics[width=\linewidth]{Figures/CS7.pdf}
%    \caption{\textcolor{red}{Two typical geographic visualisation methods depicting glacier frontal changes} }
%    \label{fig:cs-7}
%\end{figure}

\begin{figure*}[t]
    \centering
    \includegraphics[width=\textwidth]{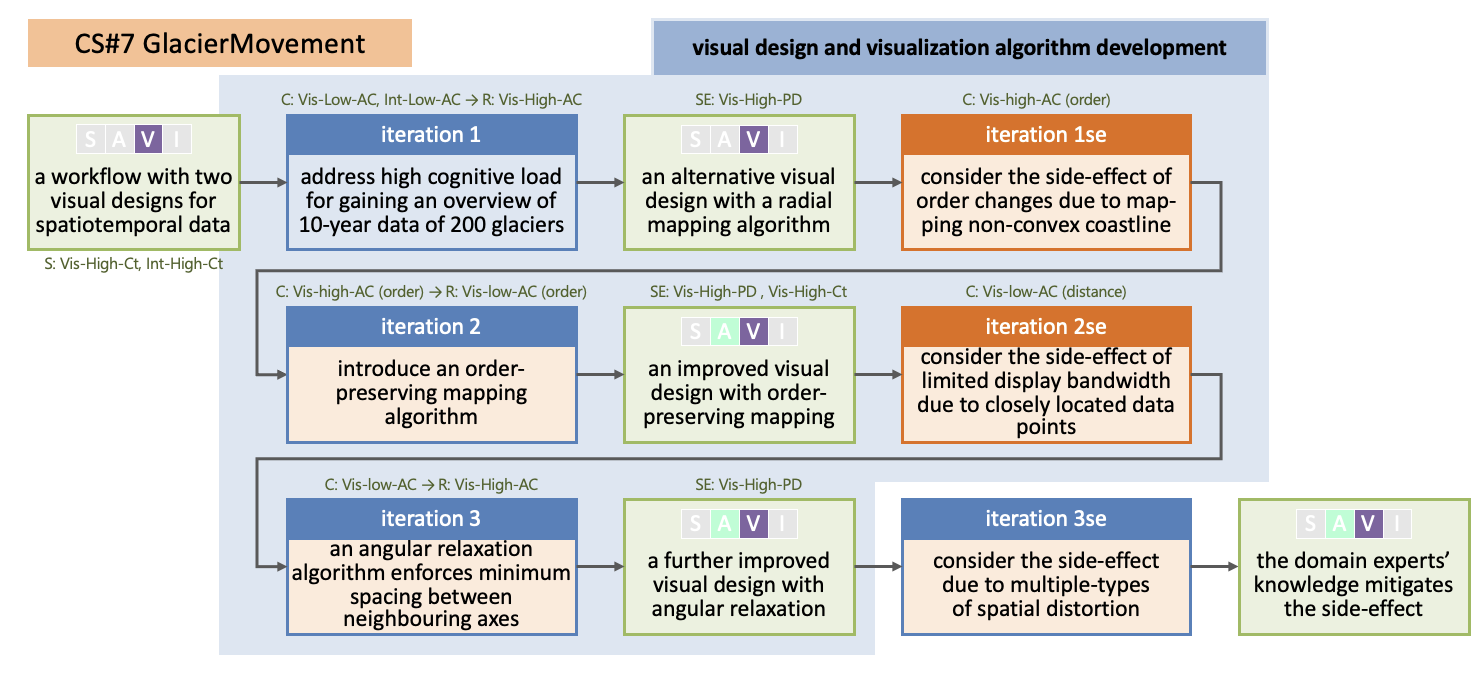}
    \caption{\textbf{CS\#7 GlacierMovement}: An updated workflow for optimizing workflows (WF4OWF) for the GlacierMovement case.}
    \label{fig:GlacierMovement}
\end{figure*}

Here, we revisit our design process retrospectively using the framework \cite{Chen:2019:CGF}. The baseline workflow exhibits a key \textbf{symptom}: scientists experience high cognitive effort when attempting to analyze both spatial and temporal patterns in glacier dynamics (Vis-High-Ct, Int-High-Ct). The underlying \textbf{cause} is that existing visualizations separate the spatial and temporal dimensions into different representations, forcing analysts to integrate the information mentally. This results in low alphabet compression and increased cognitive cost during the analytical process (Vis-Low-AC).
To address this issue, the design team introduced a \textbf{remedy} in the form of a radial visualization that integrates spatial and temporal information into a single representation (Vis-High-AC). The design takes advantage of the geographical characteristic that the glacier termini lie along the coastal boundary of Greenland. By treating the coastline as a one-dimensional boundary embedded in two-dimensional space, the visualization reduces the spatial dimension and maps glacier locations onto angular coordinates around a circle. Time is represented as concentric rings radiating outward from the center, where each ring corresponds to a different year. This radial projection allows the visualization to display nearly 200 glacier time series simultaneously while preserving their spatial ordering along the coastline. This approach makes a shift from one that suffers from Vis-Low-AC to one that increases Vis-High-AC to reduce Vis-High-Ct in the analytical process.
A naive radial mapping would distribute glaciers uniformly around a circle, thereby losing important spatial cues such as the uneven distribution of glaciers, their proximity, and their relative orientation around Greenland. This may be a \textbf{side-effect} of the \textbf{remedy} where increasing Vis-High-AC risks introducing Vis-High-PD. This is likely due to the abstraction that can affect geographic relationships. In \cite{Chen:2019:CGF} the authors highlight how increases in alphabet compression often come with increased potential distortion, this is possibly an example of this trade-off.
This introduces a new \textbf{symptom} related to loss of geographic context. The \textbf{cause} is the distortion introduced by the uniform angular mapping. To mitigate this issue, we developed a spatial mapping algorithm that preserves the natural ordering of glacier termini along the coastline. In a first step, the algorithm snaps glacier positions to the coastal boundary and then computes traversal distances between neighboring glaciers. These distances are then used to distribute the glaciers in angular space while maintaining neighborhood relationships.
Despite this improvement, glaciers spatially tightly clustered can lead to densely packed radial axes and introduce overlapping or visual clutter (Vis-High-Ct, Vis-High-PD). To address this \textbf{symptom}, we introduced an angular relaxation algorithm that enforces a minimum spacing between neighboring axes while maintaining key spatial reference points (Vis-Low-PD). These reference points correspond to glaciers located near cardinal directions (north, south, east, west), which serve as anchors for maintaining geographic orientation. Although this relaxation process improves readability and reduces clutter, it introduces a small degree of spatial distortion, representing a typical trade-off between alphabet compression and potential distortion within the framework \cite{Chen:2019:CGF}.
A further challenge concerns the effective representation of glacier advance and retreat values. Color-only encoding can make it difficult to distinguish magnitude and sign of changes, especially when galciologists compare multiple time steps (Vis-High-Ct). To overcome this limitation, the final visualization design uses divergent color schemes combined with geometric encoding. Positive values representing glacier advance are shown using blue tones, while negative values representing retreat are shown using red tones. The magnitude of the change is represented through the amplitude or thickness in radial stream-like plots. Two final visual styles were proposed: an area-based radial plot and a tube-style representation that emphasizes magnitude through thickness (reducing Vis-High-Ct, while still maintaining sufficient Vis-High-AC for overview).
The resulting visualization enables scientists to observe the behavior of all 199 glaciers over ten years within a single integrated display. Domain experts reported that visualization provided a clear overview of glacier dynamics and helped identify large retreat events and regional patterns that were previously difficult to detect. The visualization also helped identify anomalies in the processed data, demonstrating its usefulness not only for analysis but also for data validation.
This case study illustrates how a visualization design approach can optimize a scientific workflow by improving the balance between alphabet compression, potential distortion, and cognitive cost. By exploiting the boundary-based spatial structure of the dataset and integrating spatial and temporal information into a radial representation, the visualization significantly improves the efficiency of glacier behavior analysis compared to traditional methods. The Workflow for Optimizing Workflows (WF4OWF) for the GlacierMovement case is depicted in \textit{Figure} \ref{fig:GlacierMovement}.

%% file: Appendix/AppendixH.tex
\begin{table*}[t]
\centering
\caption{A summary of case studies CS\#A $\sim$ CS\#G and their main characteristics.}
\label{tab:case-studies-2}
\renewcommand{\arraystretch}{1.35}
\setlength{\tabcolsep}{3pt}
\scalebox{0.78}{%
\begin{tabular}{|
  >{\raggedright\arraybackslash}p{2.6cm}|
  >{\raggedright\arraybackslash}p{2.6cm}|
  >{\raggedright\arraybackslash}p{2.6cm}|
  >{\raggedright\arraybackslash}p{2.6cm}|
  >{\raggedright\arraybackslash}p{2.6cm}|
  >{\raggedright\arraybackslash}p{2.6cm}|
  >{\raggedright\arraybackslash}p{2.6cm}|
  >{\raggedright\arraybackslash}p{2.6cm}|}
\hline

 & \textbf{CS\#A}
 & \textbf{CS\#B}
 & \textbf{CS\#C}
 & \textbf{CS\#D}
 & \textbf{CS\#E}
 & \textbf{CS\#F}
 & \textbf{CS\#G} \\
 \textbf{Characteristics}
 & \textbf{DashboardDesign}
 & \textbf{RadialIcicleTree}
 & \textbf{SensitivityAnalysis}
 & \textbf{MusicMood}
 & \textbf{MicroblogData}
 & \textbf{TopoText}
 & \textbf{MetricsVis}
\\ \hline
 
\textbf{Application Context}
 & Public health (COVID-19 dashboard design)
 & Tree visualization (visual design)
 & Epidemiological modeling (sensitivity analysis)
 & Music / ML (ensemble predictions of music mood)
 & Emergency management (microblog data analysis)
 & Geospatial text exploration (context-preserving text data)
 & Law enforcement (performance evaluation of officers)
\\ \hline
 
\textbf{User}
 & Dashboard users (e.g., healthcare managers)
 & VIS designers / users of tree data
 & Epidemiologists, modeling scientists
 & Music experts, ML developers
 & Causal analysis experts in emergency / law enforcement
 & Analysts of geospatial text data
 & Supervisors in law enforcement agencies
\\ \hline
 
\textbf{Type of Resultant Workflow}
 & Human-centric process
 & Human-centric process
 & Hybrid of human- \& machine-centric
 & Hybrid of human- \& machine-centric
 & Hybrid of human- \& machine-centric
 & Human-centric process
 & Hybrid of human- \& machine-centric
\\ \hline

\textbf{Visualization in Workflows}
 & Initial: with Vis \newline Final: with Vis
 & Initial: with Vis \newline Final: with Vis
 & Initial: with Vis \newline Final: with Vis
 & Initial: no Vis \newline Final: with Vis
 & Initial: with Vis \newline Final: with Vis
 & Initial: with Vis \newline Final: with Vis
 & Initial: no Vis \newline Final: with Vis
\\ \hline

\textbf{Remedy Types} \newline (Stat, Alg, Int, Vis)
 & Stat:~0, Alg:~0, \newline Int:~1, Vis:~4
 & Stat:~0, Alg:~2, \newline Int:~0, Vis:~4
 & Stat:~0, Alg:~2, \newline Int:~1, Vis:~2
 & Stat:~0, Alg:~0, \newline Int:~1, Vis:~2
 & Stat:~0, Alg:~1, \newline Int:~1, Vis:~1
 & Stat:~0, Alg:~0, \newline Int:~0, Vis:~7
 & Stat:~1, Alg:~1, \newline Int:~1, Vis:~1
\\ \hline

\textbf{Analysis Mode}
 & Prospective
 & Prospective
 & Prospective \& Retrospective
 & Prospective
 & Retrospective
 & Retrospective
 & Retrospective
\\ \hline
 
\textbf{Optimization Workflow Mode}
 & Iteration + Branching
 & Iteration
 & Branching
 & Iteration
 & Iteration
 & Branching + Iteration
 & Iteration
\\ \hline
 
\textbf{Number of Opt. Sub-workflows}
 & ${\sim}$4
 & ${\geq}$6
 & ${\sim}$3
 & ${\sim}$3
 & ${\sim}$3
 & 3
 & ${\sim}$2
\\ \hline
 
%\textbf{Research Background of Domain Expert}
% & Scientific visualization, data science, information theory, machine learning
% & Scientific visualization, data science, information theory, machine learning
% & Scientific visualization, data science, information theory, machine learning
% & Scientific visualization, data science, information theory, machine learning
% & Scientific visualization, data science, information theory, machine learning
% & Scientific visualization, data science, information theory, machine learning
% & Scientific visualization, data science, information theory, machine learning
% \\ \hline
 
\end{tabular}%
}
\end{table*}

% ====================
\section{Summary of Past Case Studies}
\label{appendix:H}

In this appendix, we summarize our observations of the seven existing case studies conducted before 2025. During our action research project, we applied the attributes that were used to observe the case studies in the project to these existing case studies. Our observations are briefly summarized in Table \ref{tab:case-studies-2}.
The reports of the seven existing case studies can be found on the IVAS web site \cite{IVAS:2026:web}. They are:

\begin{itemize}
    \item \textbf{CS\#A Dashboard} -- A prospective analysis of why many dashboards during the COVID-19 pandemic displayed numbers in large fonts as part of the discussion if this was against the VIS design principles: (i) overview first and details on demand, and (ii) encode data visually \cite{Bach:2023:TVCG}.
    \item \textbf{CS\#B RadialIcicleTree} -- A prospective analysis of how two commonly-used visual designs could be improved with the development of a new visual design called Radial Icicle Tree \cite{Jin:2024:TVCG}.
    \item \textbf{CS\#C SensitivityAnalysis} -- A primarily prospective analysis of how a workflow for sensitivity analysis in epidemiological modeling could be improved in two different ways \cite{Rydow:2023:TVCG}. One solution was first proposed and then analyzed in abstraction, and another was analyzed in abstraction first, and the abstract remedy was then instantiated.
    \item \textbf{CS\#D MusicMood} -- A prospective analysis of how a commonly-used visual design could be improved and how an ML workflow for music mood classification could be improved \cite{Ye:2023:TVCG}.
    \item \textbf{CS\#E MicroblogData} -- A retrospective analysis of how a workflow for analyzing microblog data in the context of emergency and law enforcement was improved \cite{Zhang:2016:CGF}. It is one of the three
    case studies reported in the original methodology paper \cite{Chen:2019:CGF}.
    \item \textbf{CS\#F TopoText} -- A retrospective analysis of three visual designs that were compared in an empirical study. It is one of the three case studies reported in the original methodology paper \cite{Chen:2019:CGF}.
    \item \textbf{CS\#G MetricVis} -- A retrospective analysis of how a workflow for performance evaluation in the context of law enforcement was improved \cite{Zhao:2017:THS}. It is one of the three
    case studies reported in the original methodology paper \cite{Chen:2019:CGF}.
\end{itemize}

The rows have the following characteristics:

\begin{itemize}
    \item \textbf{Application Context} -- The application context of the workflow to be optimized.
    \item \textbf{User} -- The target users of the workflow to be optimized.
    \item \textbf{Type of Resultant Workflow} -- The types of major processes in the workflow after improvement. Options are: human-centric | machine-centric | both.
    \item \textbf{Visualization in Workflows} -- Comparison of the visualization components in the baseline workflow and the resultant workflow.
    \item \textbf{Remedy Type} -- The number of VA components featured in the remedy. Four types of components are: statistics, algorithms, visualization, and interaction.
    \item \textbf{Analysis Mode} -- The operational mode of the SCORE analysis. Options are: retrospective | prospective.
    \item \textbf{Optimization Workflow Mode} -- The structural features of the workflow for optimizing VA workflows. Options are: Iteration | Nesting | Branching.
    \item \textbf{Number of Opt. Sub-workflows} -- The number of sub-workflows in the optimization workflow (note: not the workflow to be optimized).
\end{itemize}